\documentclass[iop]{emulateapj}
\usepackage{epsfig, natbib} 
\usepackage{mathrsfs,amssymb}
\usepackage{graphicx,latexsym}
\newcommand{\Ha}{H$\alpha$}
\newcommand{\hii}{{\sc Hii}}
\newcommand{\hi}{{\sc Hi}}

\newcommand{\oiii}{{[\sc Oiii]}}
\newcommand{\sii}{{[\sc Sii]}}

\newcommand{\vsini}{$v \sin{i}$}
\newcommand{\kms}{\rm km\ s^{-1}}

\def\spose#1{\hbox to 0pt{#1\hss}}
\def\dt{\spose{\raise 1.0ex\hbox{\hskip2pt$\mathchar"201$}}}    

\slugcomment{Accepted to the Astrophysical Journal March 10, 2013}

\shorttitle{In-Situ Field OB Stars}
\shortauthors{Oey et al.}

\begin{document}

\title{A Sample of OB Stars That Formed in the Field}

\author{M. S. Oey, J. B. Lamb, C. T. Kushner, E. W. Pellegrini\altaffilmark{1}, and
  A. S. Graus\altaffilmark{2}} 
\affiliation{Department of Astronomy, University of Michigan, 830 Dennison Building, 500 Church Street, Ann Arbor, MI, 48109-1042}

\altaffiltext{1}{Present address:  Department of Physics and Astronomy, University of
  Toledo, 2801 W. Bancroft, Toledo, OH 43606 }
\altaffiltext{2}{Present address:  Department of Physics and Astronomy, University of
  California, Irvine, CA 92697}


\begin{abstract}

We present a sample of 14 OB stars in the Small Magellanic Cloud that
meet strong criteria for having formed under extremely sparse star-forming
conditions in the field.  These stars are a minimum of 28 pc in
projection from other OB stars, and they are centered within
symmetric, round \hii\ regions.  They show no evidence of bow shocks,
implying that the targets are not transverse runaway stars.  Their
radial velocities relative to local \hi\ also indicate that they are
not line-of-sight runaway stars.  A friends-of-friends analysis shows
that 9 of the objects present a few low-mass companion
stars, with typical mass ratios for the two highest-mass stars of around
0.1.  This further substantiates that these OB stars formed in place,
and that they can and do form in extremely sparse conditions.  This
poses strong constraints on theories of star formation and challenges
proposed relations between cluster mass and maximum stellar mass.

\end{abstract}

\section{Introduction} 

One of the most important constraints on theories of massive star
formation is the degree to which massive, OB stars can form in
true isolation (see discussion in, e.g., Lamb et al. 2010).  Our
conventional understanding of the stellar initial 
mass function (IMF) dictates that a large population of lower-mass
stars should generally accompany the formation of massive stars.
Similarly, the initial cluster mass function can be described as a
clustering law where the number of massive stars $N_*$ is described by
an inverse power-law function almost identical to the IMF (e.g., Oey 2011;
Oey \& Clarke 1998; Elmegreen \& Efremov 1997).  Both the IMF and
initial cluster mass function behave as probability density
functions, so if they are invariant, it
implies that individual OB stars should form in isolation
on rare occasions.  But does Nature actually operate that way?

Isolated field OB stars are certainly known to exist.  However,
whether any truly isolated OB stars actually formed {\it in situ} is
presently controversial.  A substantial fraction of the field massive
star population is known to consist of runaway stars ejected from clusters
but the contribution, if any, of an {\it in-situ} component is not
established.  However, several lines of evidence support the
formation of O stars in extremely sparse groups, if not in isolation
(Oey \& Lamb 2012):  (1) The number of field O stars defined by a
friends-of-friends algorithm falls smoothly onto the power-law $N_*$
distribution for groups and clusters (Oey et al. 2004).  This implies
that the field star population simply represents the low-mass extreme
for the mass spectrum of clustered star formation.  While thus far this has
been evaluated only for the Small Magellanic Cloud (SMC), the star
formation properties in this galaxy are unremarkable, and the relation
most likely applies in other star-forming galaxies as well.  (2) The
IMF for massive field stars is substantially steeper than for clusters
and associations in the Magellanic Clouds (Lamb et al. 2013; Massey
2002; Massey et al. 1995), 
having a power-law slope $\Gamma\lesssim -2.3$ in contrast to the Salpeter
(1955) value of --1.35.  This is perhaps the only systematic variation in the
IMF that is now well established.  Since the frequency of runaway
stars {\it increases} with mass (e.g., Moffat 1998; Gies 1987), which
would flatten the inferred IMF, the steep field
IMF suggests a major, if not dominant, contribution of {\it in situ} stars.
%
(3) The distribution of rotation velocities for field early-type stars
is also well established to differ from that for cluster members
(e.g., Guthrie 1984; Wolff et al. 2007), with the field \vsini\ 
strongly weighted to low values relative to clusters.  This may be an
evolutionary spin-down effect (Huang \& Gies 2008), or it may be
intrinsic to the star-formation density (Wolff et al. 2007).
(4) Last, but not least, there are numerous
examples of candidate {\it in situ} massive field stars.  Lamb et
al. (2010) and de Wit et al. (2004, 2005) identify OB stars with
sparse associated groups of lower-mass stars, suggesting that
a significant fraction of massive stars form in such extreme,
low-density environments.  Similarly, Testi et al. (1997, 1999)
identify isolated Herbig Ae/Be stars.  In that case in particular, it
is difficult to reconcile the isolation of these stars with their youth.  

\begin{deluxetable*}{clcccccccccc}
\scriptsize
\tablewidth{0pt}
\tablecolumns{12}
\tablecaption{SMC in-situ field OB stars}
\tablehead{\colhead{Star\tablenotemark{a}} & \colhead{Sp Type\tablenotemark{b}} & 
\colhead{RA and Dec (J2000)\tablenotemark{c}} & \colhead{RV\tablenotemark{d}}
 & \colhead{$v$(\hi)\tablenotemark{d}} & \colhead{$V$\tablenotemark{c}}
& \colhead{$A_V$\tablenotemark{e}} & \colhead{$l$ (pc)} &
\colhead{$N$(tot)} & \colhead{$N$(MS)} &
\colhead{$m_1\ (M_\odot)$\tablenotemark{b}} & \colhead{$m_2/m_1$}
}
\startdata
1600 & O8.5 V & 0:42:09.92 ~ --73:13:56.9 & ~93 & 103 & 14.725 & 0.42 & 0.98 & 3 & 2 & 23.6 & 0.11 \\
3173 & O$+$neb & 0:43:36.69 ~ --73:02:26.9 & 111 & 111 & 15.529 & 0.37 & 1.06 & 5 & 3 & 25--40 & $\leq 0.2$ \\
17813 & B0 V & 0:50:49.99 ~ --73:24:22.3 & 192 & 184 & 15.101 & 0.38 & 0.91 & 0 & 0 & 18.3 & \nodata \\
24119 & B1--2.5 V & 0:52:38.19 ~ --73:26:17.1 & 170 & 170 & 15.272 & 0.27 & 0.91 & 4 & 3 & 14.8 & 0.16 \\
35491 & O8 V & 0:55:59.59 ~ --72:19:54.4 & 126 & 126 & 14.856 & 0.47 & 0.83 & 0 & 0 & 30.4 & \nodata \\
36514 & O9 V & 0:56:17.33 ~ --72:17:28.8 & 141 & 141 & 15.385 & 0.33 & 0.98 & 12 & 7 & 20.5 & 0.12 \\
64453 & B1 Ib & 1:06:40.30 ~ --73:10:24.6 & 194 & 194 & 13.506 & 0.29 & 1.59 & 2 & 2 & 19.0 & 0.12 \\
66415 & O9.7 Ia & 1:07:40.40 ~ --72:51:00.0 & 185 & 185 & 13.285 & 0.26 & 1.44 & 4 & 2 & 34.5 & 0.08 \\
67334 & B0 V & 1:08:08.03 ~ --72:38:19.7 & 173 & 173 & 15.270 & 0.43 & 1.44 & 4 & 1 & 16.5 & 0.19 \\
69598 & O9 V & 1:09:26.78 ~ --72:01:26.9 & 164 & 168 & 15.640 & 0.23 & 2.04 & 11 & 6 & 17.6 & 0.51 \\
70149 & O9.5 V & 1:09:48.26 ~ --72:30:19.4 & 117 & 126 & 14.755 & 0.31 & 1.51 & 3 & 0 & 23.8 & \nodata \\
71409 & Be & 1:10:45.17 ~ --72:21:37.5 & 184 & 184 & 15.029 & 0.25 & 1.82 & 0 & 0 & 15--20 & \nodata \\
71815 & O8 V & 1:11:05.63 ~ --72:13:41.7 & 172 & 172 & 15.495 & 0.18 & 1.97 & 0 & 0 & 19.1 & \nodata \\
75984 & B0.2 V & 1:15:14.70 ~ --72:20:19.4 & 216 & 213\tablenotemark{f} & 15.693 & 0.31 & 2.49 & 11 & 7 & 14.0 & 0.36
\enddata
\tablenotetext{a}{From Massey (2002).}
\tablenotetext{b}{From Lamb et al. (2013).}
\tablenotetext{c}{From OGLE III (Udalski et al. 2008).}
\tablenotetext{d}{Velocity in $\kms$ of the \hi\ component having
  brightness temperature $\geq20$ K, that is nearest to the stellar RV.}
\tablenotetext{e}{From Zaritsky et al. (2002).}
\tablenotetext{f}{\hi\ is only observed for velocities $<215\ \kms$, so this value
  is most likely a lower limit.}
\label{t_sample} 
\end{deluxetable*}

Together, these arguments strongly suggest that massive stars
occasionally do form in relative isolation.  It was already suggested
decades ago that high-mass stars may sometimes form in the
lower-density cluster outskirts (Burki 1978).  However, since clusters may 
disperse quickly, especially when a single massive star is present
(e.g., Allison et al. 2009), it has also been suggested that
essentially all massive field stars are runaways from clusters (e.g.,
Gvaramadze et al. 2012).  The strongest evidence for the in-situ
formation of field massive stars
is to identify newborn, isolated massive stars.  The Herbig Ae/Be
stars of Testi et al. are an important such data set.  More recently,
Selier et al. (2011) identify the SMC object N33 as an isolated,
compact \hii\ region hosting a candidate isolated field O star, and
Bressert et al. (2012) present a candidate sample of relatively
isolated O stars near the 30 Doradus giant star-forming complex.
We also note a number of spherical, Str\"omgren sphere \hii\ regions
with centrally positioned stars
identified by Zastrow et al. (2013) in the Large Magellanic Cloud.
In this contribution, we present a substantial sample of 14 field OB
stars in the SMC hosting single-star \hii\ regions that appear to
imply the {\it in-situ} formation of these stars.

\section{SMC Field Stars in Str\"omgren Spheres}

Oey et al. (2004) determined the SMC clustering law for
massive stars by applying a friends-of-friends algorithm to
photometrically identified OB star candidates.  Any stars not
within the 28-pc clustering length from other candidates are
defined to be field stars, thereby yielding an essentially complete
sample of field stars in this galaxy.  
We carried out a complete spectroscopic survey of this field star
sample, the Runaways and Isolated O-Type Star Spectroscopic Survey of
the SMC (RIOTS4), using the IMACS imaging spectrograph at Magellan
with multislit masks (Lamb et al. 2013; Oey \& Lamb 2012).  This
survey yields, among other things, spectral types and radial
velocities of the target stars.  Critical insight on the origin of
these stars can be provided by the \hii\ regions that they generate,
which can constrain the likelihood that they formed in the
observed locations.  Fortuitously, we recently completed a new
catalog of \hii\ regions in the Magellanic Clouds (Pellegrini et
al. 2012, 2013) using the \Ha,
\oiii, and \sii\ narrow-band imaging data from the Magellanic Clouds
Emission-Line Survey (MCELS; Smith et al. 2005).  

We are thus in a position to carefully examine the nebulae for this
sample of SMC field OB stars that have known spectral types and radial
velocities.  In future work, we
will systematically address the quantitative and varied properties of
the entire field star sample, but the subset we present here provides
especially compelling examples of objects that imply that the host
stars were formed in place.  We selected \hii\ regions that appear the most
symmetric, with no evidence of bow shocks, and with target stars that are
well centered within these nebulae.  We examine these objects in both
\Ha\ and in the ratio map of \oiii/\sii, since the ionization
structure confirms the general symmetry and allows us to identify
objects with high optical depth (Pellegrini et al. 2012).  In
addition, we limit the sample to 
stars which show strong isolation with minimal evidence of companions. 

Table~\ref{t_sample} lists the sample.  Columns 1, 2 and 3
respectively give the star's Massey (2002) identification, RIOTS4 spectral
type from Lamb et al. (2013), and position from OGLE III (Udalski et
al. 2008).  Columns 4
and 5 list our RIOTS4 stellar radial velocity and the radial velocity
of the nearest significant \hi\ component from Stanimirovi\' c et
al. (1999), as described below.  The OGLE $V$ magnitude and
extinction $A_V$ from Zaritsky et al. (2002) are given in columns 6
and 7, respectively.  Columns 8 -- 12 list clustering parameters
for each star described in \S 3 below.

\begin{deluxetable}{rlrccr}
\scriptsize
\tablewidth{0pt}
\tablecolumns{6}
\tablecaption{\hii\ Region Properties}
\tablehead{
\colhead{Star} & \colhead{MCELS} & \colhead{$R$(arcsec)} & 
\colhead{$F$(\Ha)\tablenotemark{a}} & \colhead{SB\tablenotemark{b}} & 
\colhead{$n_e(\rm cm^{-3}$)}
}
\startdata
1600  & S6  &  156.8  & 1.08E-11 & 1.39E-16 & 0.9 \\
3173  & S14  &  12.0  & 4.35E-12 & 9.61E-15 & 25.9 \\
17813 & S73  &  29.0  & 1.45E-12 & 5.52E-16 & 4.0 \\
24119 & S88  &  26.0  & 2.92E-12 & 1.39E-15 & 6.7 \\
35491 & \nodata  &  75.5  & 7.88E-12 & 4.40E-16 & 2.2 \\
36514 & S110  &  35.8  & 5.57E-12 & 1.38E-15 & 5.7 \\
64453 & S165  &  67.8  & 1.05E-12 & 7.35E-17 & 1.0 \\
66415 & S168  &  208.7  & 1.25E-11 & 9.76E-17 & 0.6 \\
67334 & \nodata  &  56.0  & 2.57E-12 & 2.14E-14 & 17.9 \\
69598 & S172  &  101.0  & 2.88E-12 & 9.31E-17 & 0.9 \\
70149 & S174  &  110.2  & 3.78E-12 & 9.95E-17 & 0.9 \\
71409 & S175  &  45.2  & 8.50E-13 & 1.37E-16 & 1.6 \\
71815 & S177  &  120.0  & 4.09E-12 & 9.04E-17 & 0.8 \\
75984 & S194  &  19.0  & 6.12E-13  & 5.43E-16 & 4.9
\enddata
\tablenotetext{a}{$\rm erg\ s^{-1} cm^{-2}$}
\tablenotetext{b}{$\rm erg\ s^{-1} cm^{-2} arcsec^{-2}$}
\label{t_nebulae}
\end{deluxetable}

Table~\ref{t_nebulae} gives the properties of the \hii\ regions associated
with the sample stars.  Columns 1 and 2 list the star ID as before,
and the nebular MCELS ID from Pellegrini et al. (2013).  The nebular
radius $R$, listed in column 3, is the mean of the nebular major and
minor axes.  The remaining columns give the \Ha\ flux $F$(\Ha),
\Ha\ surface brightness (SB), and the nebular electron density $n_e$,
derived from the emission measure implied by the SB and $R$.
The surface brightnesses and electron densities of these objects
in Table~\ref{t_nebulae} are commensurate with those of typical
\hii\ regions, and are an order of magnitude greater than the diffuse,
warm, ionized component of the interstellar medium (ISM), where $n_e$
is on the order of $\sim 0.1\ \rm cm^{-3}$. 
We assume an SMC distance of 60 kpc (e.g., Harries et al. 2003).

\begin{figure*}
\plotone{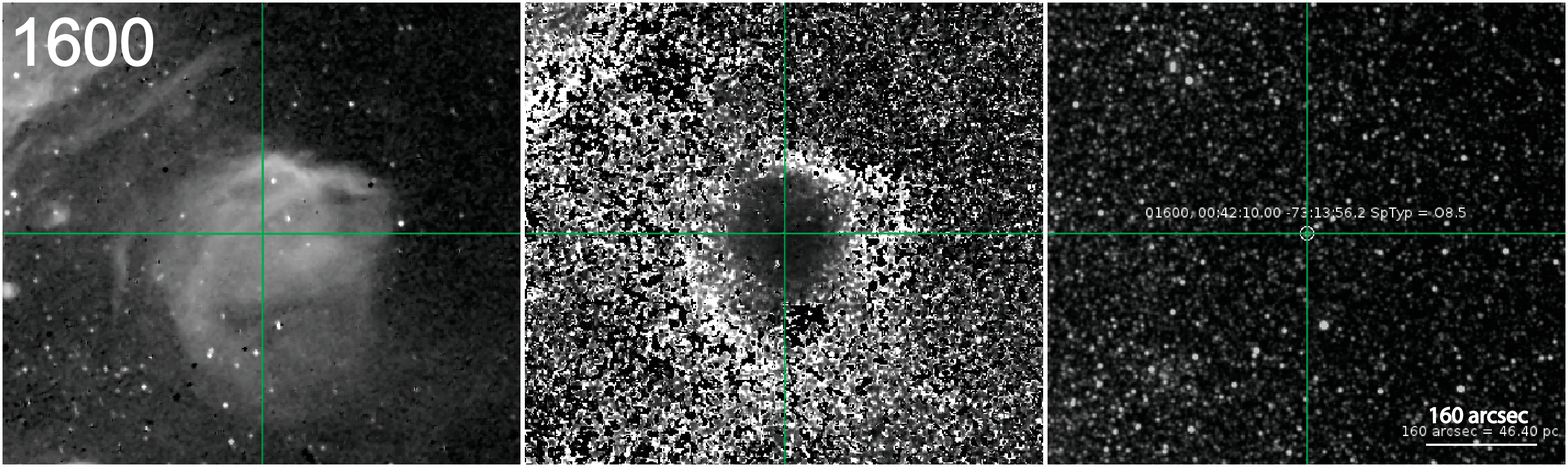}
\plotone{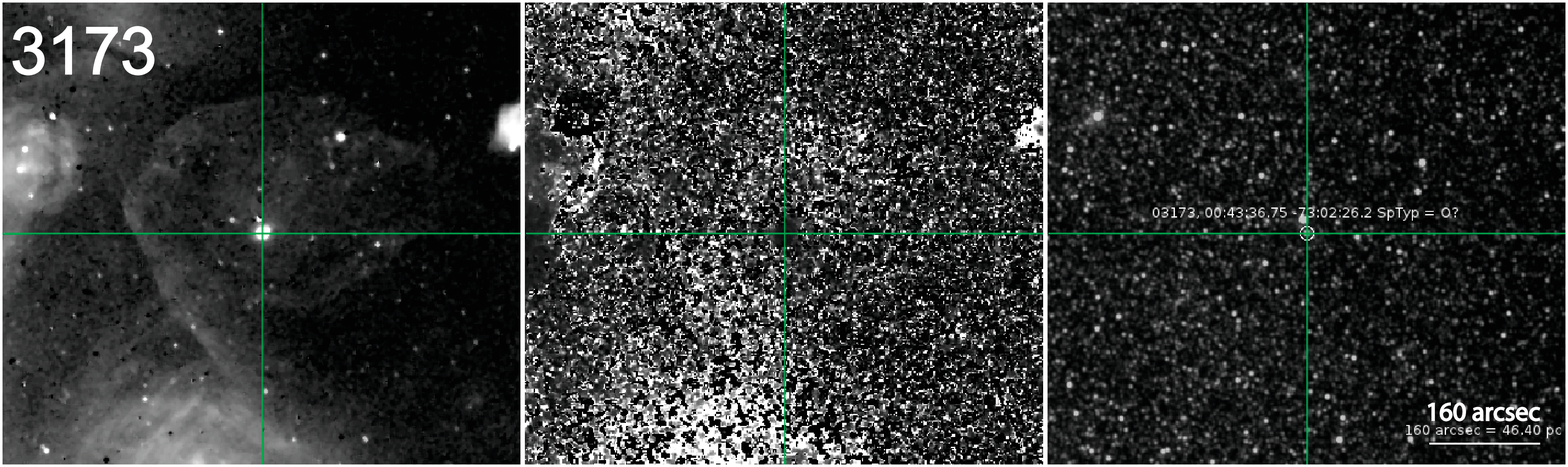}
\plotone{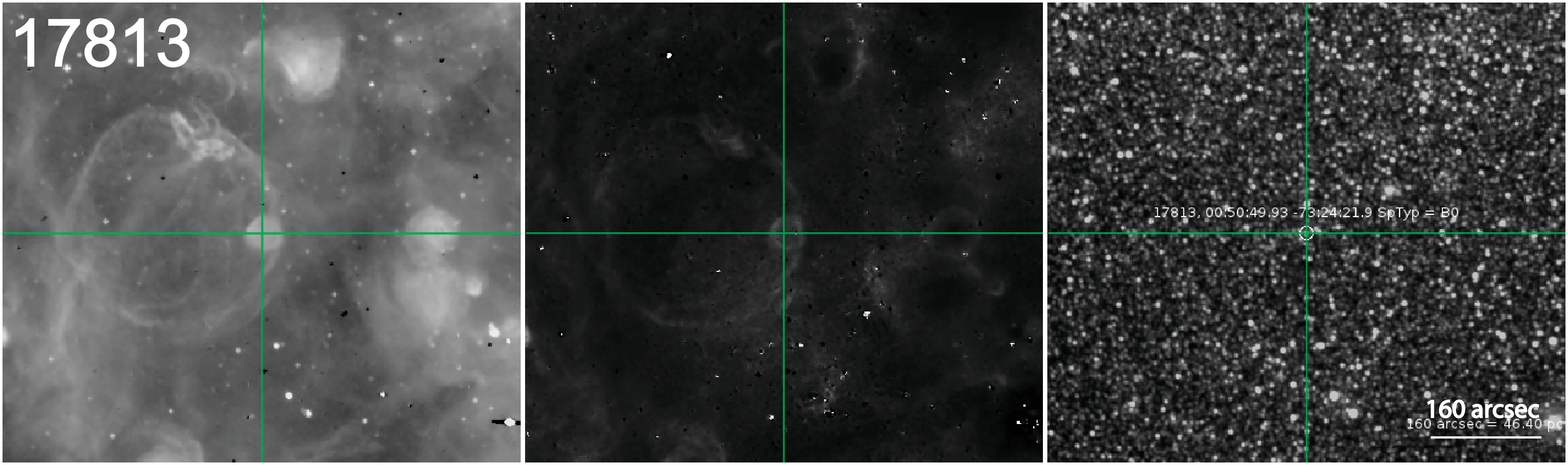}
\plotone{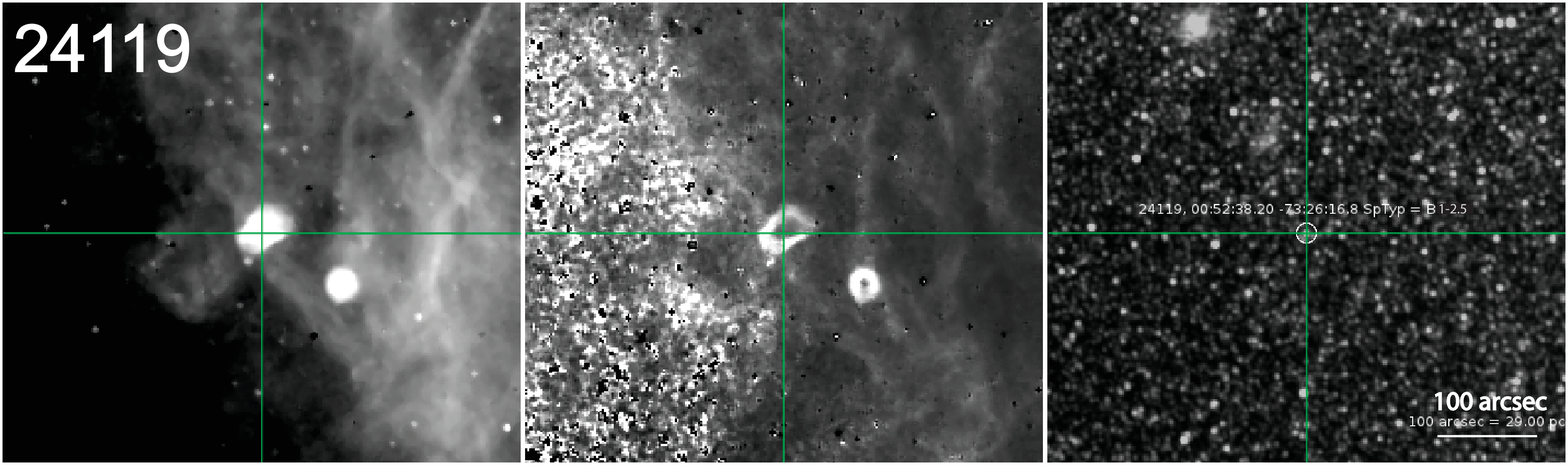}
\plotone{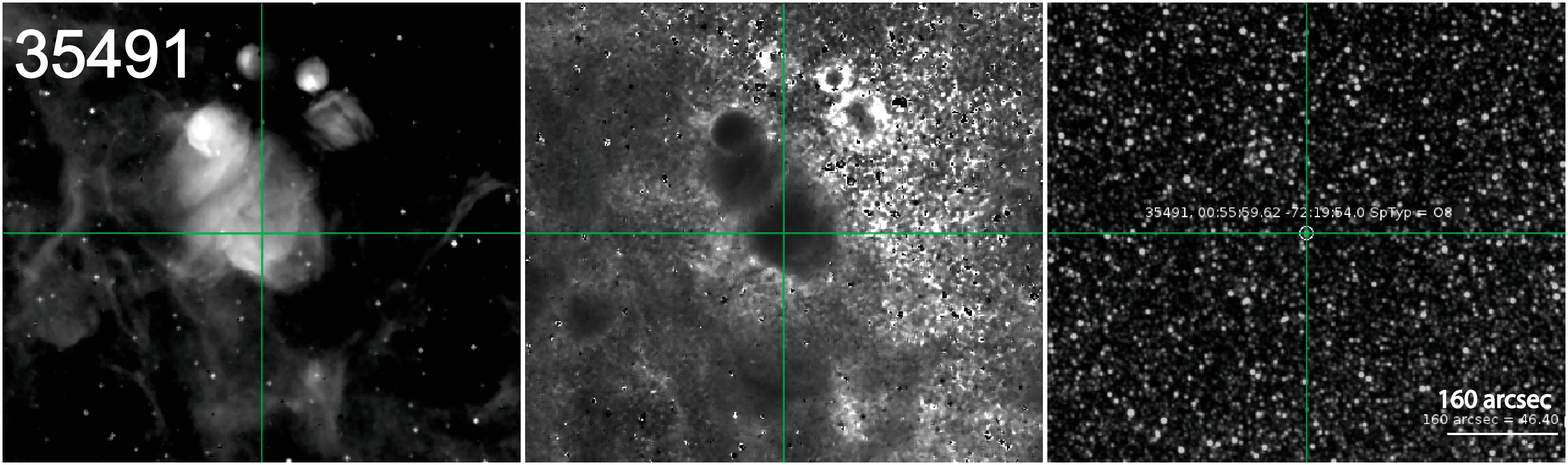}
\caption{
Images in (left to right) \Ha, \sii/\oiii, and $\lambda$5130
continuum, with the cross hairs centered on the target stars.
Strong \sii\ is black in the center panels.  The 160\arcsec\ and 100
\arcsec\ scale bars correpond to 46.4 pc and 29.0 pc at the SMC
distance, respectively.  N is up and E to the left.
\label{f_sample}}
\end{figure*}

\setcounter{figure}{0}
\begin{figure*}
\plotone{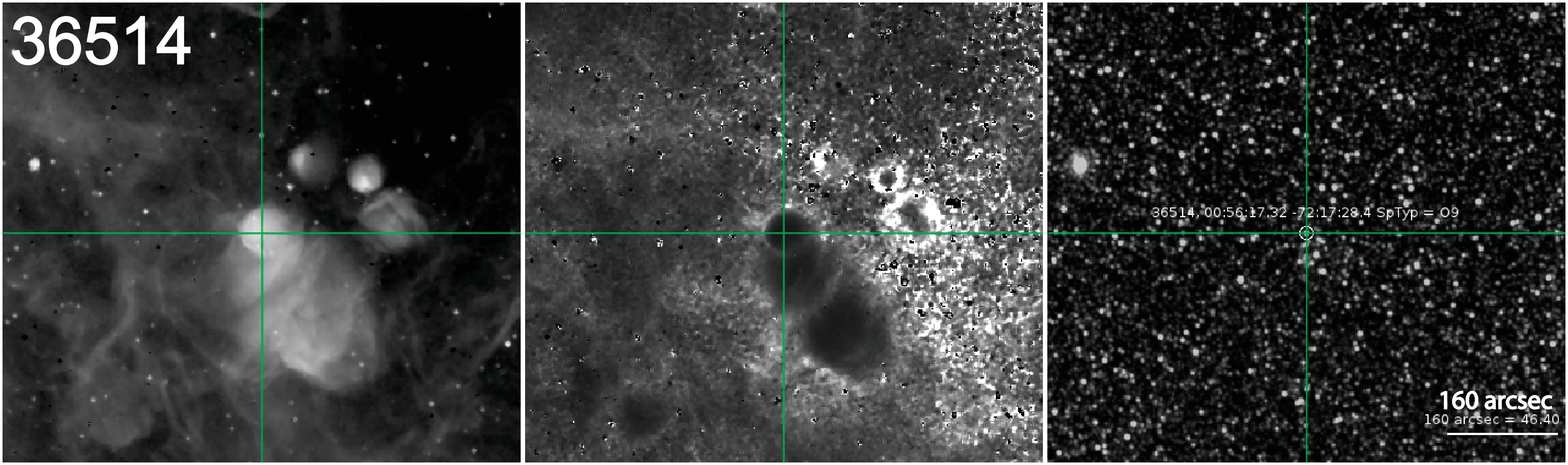}
\plotone{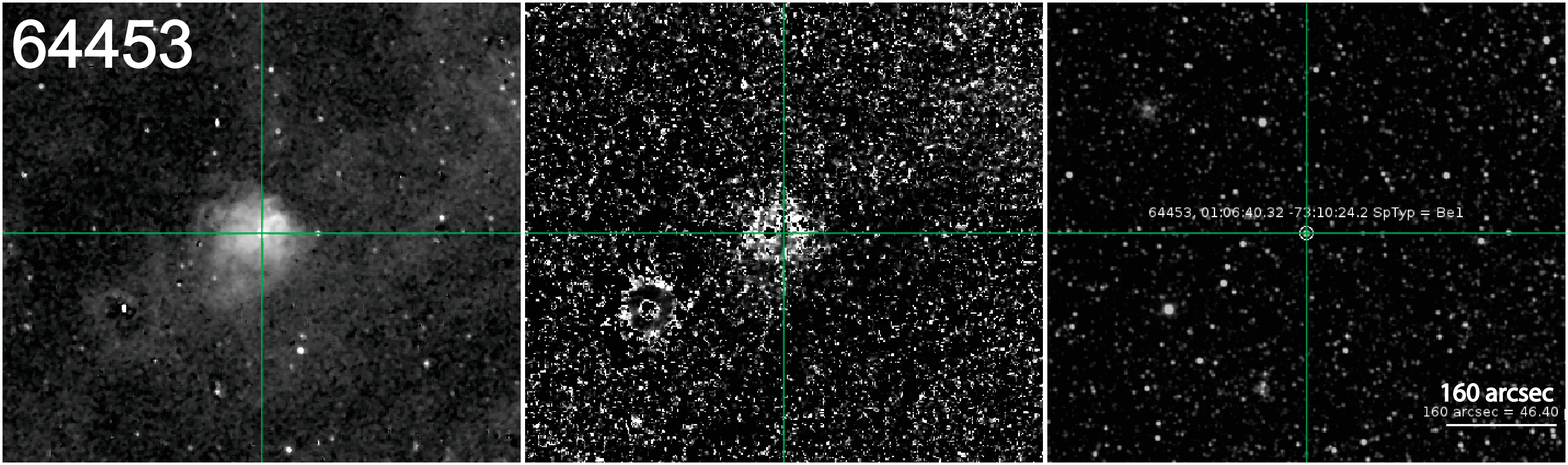}
\plotone{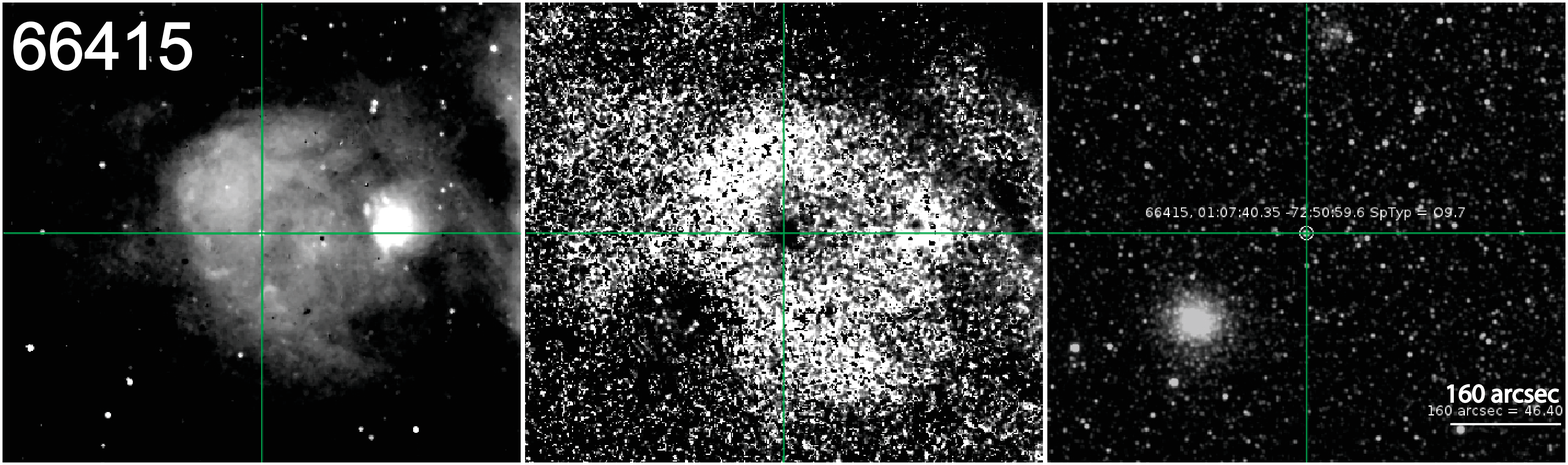}
\plotone{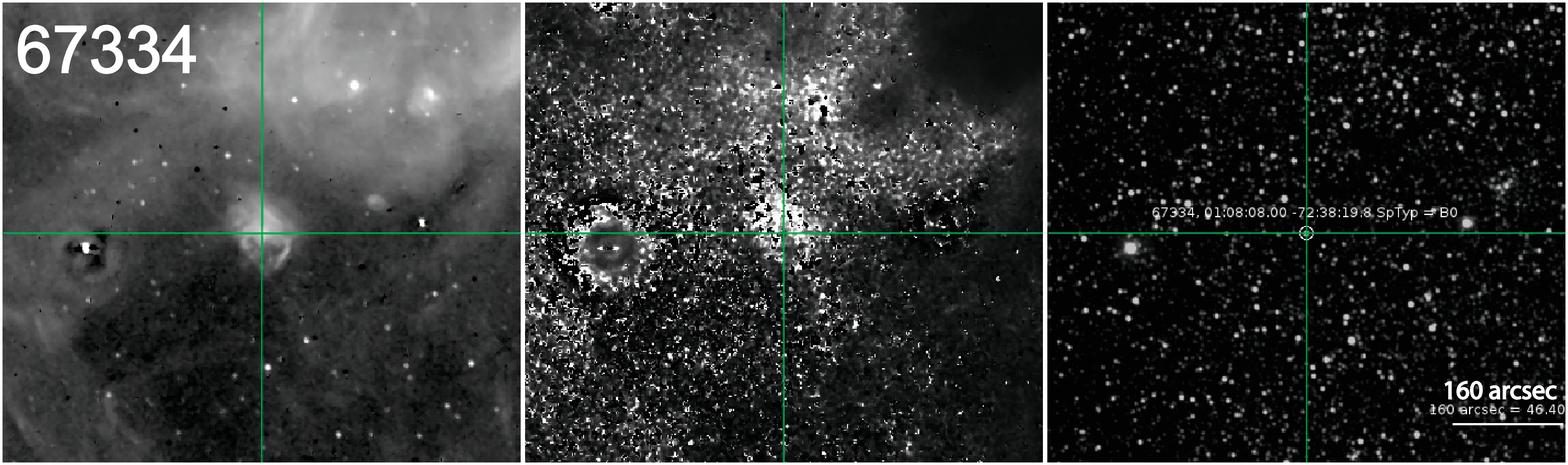}
\plotone{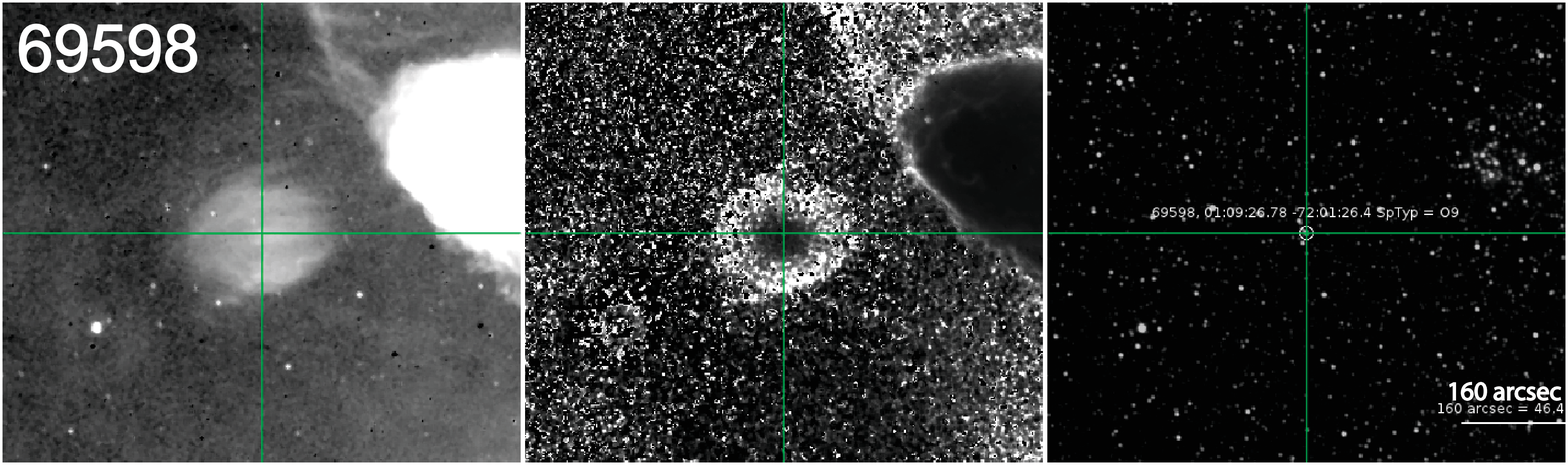}
\caption{
--- continued.}
\end{figure*}

\setcounter{figure}{0}
\begin{figure*}
\plotone{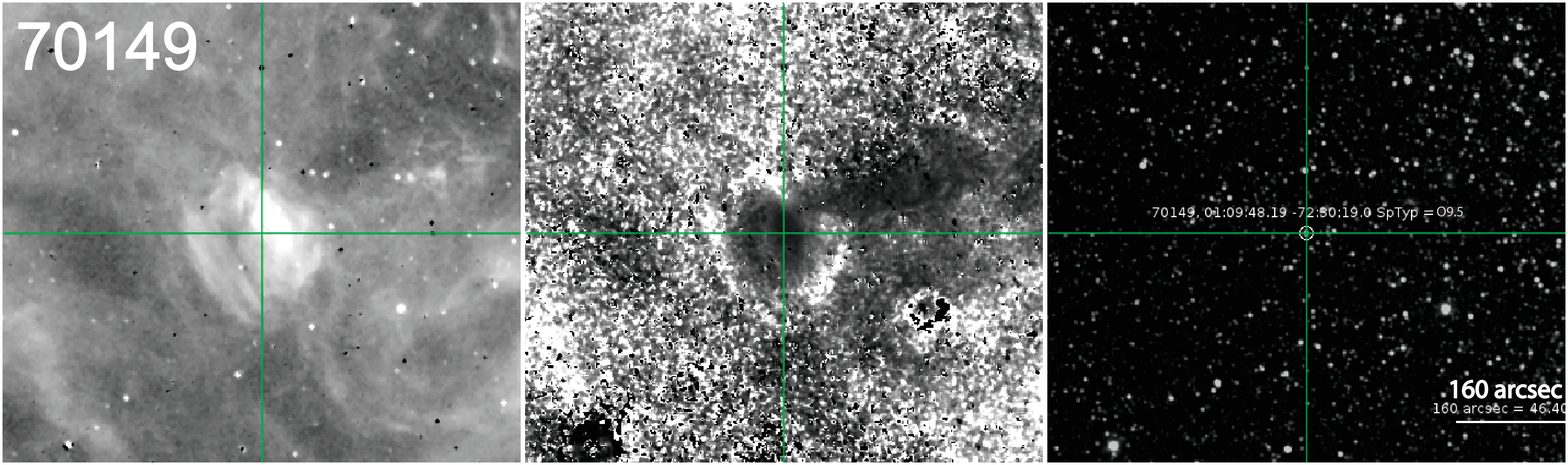}
\plotone{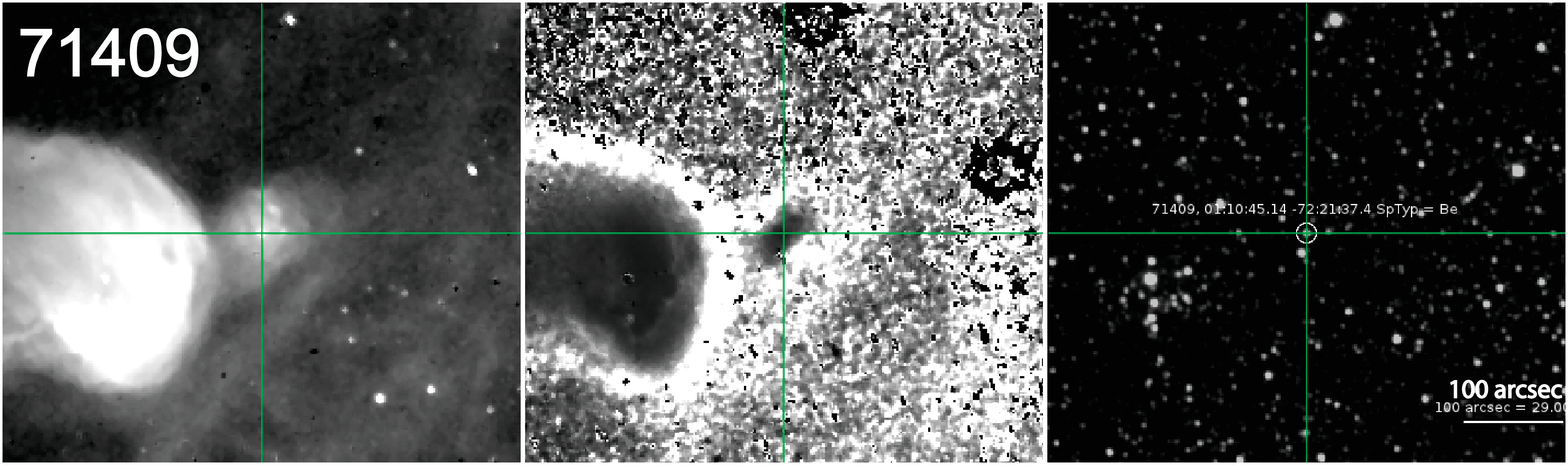}
\plotone{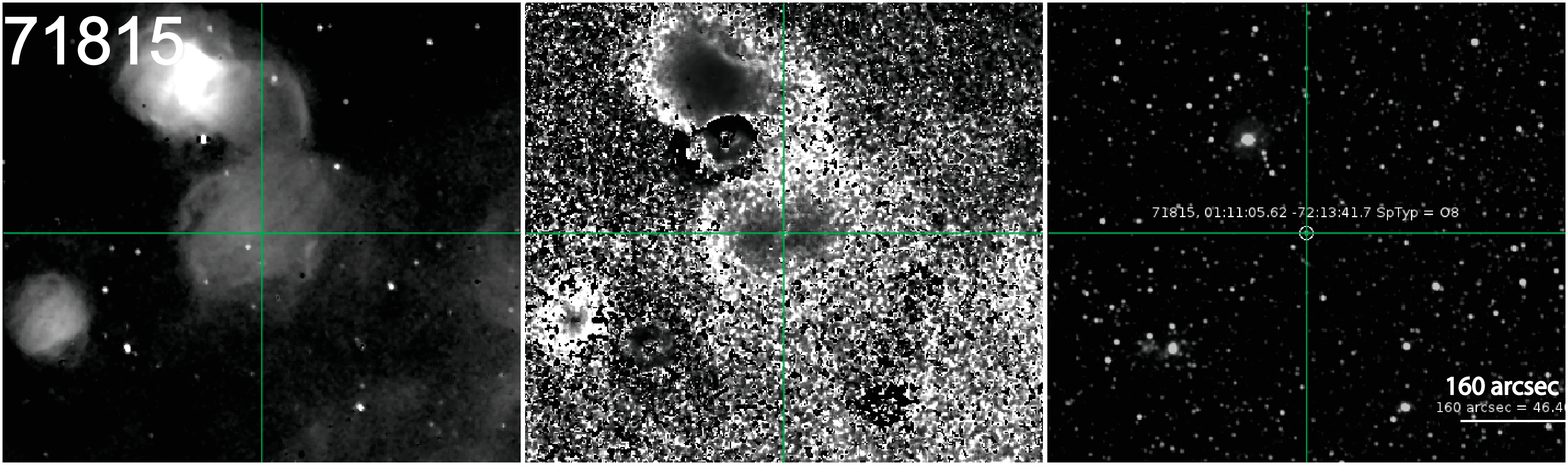}
\plotone{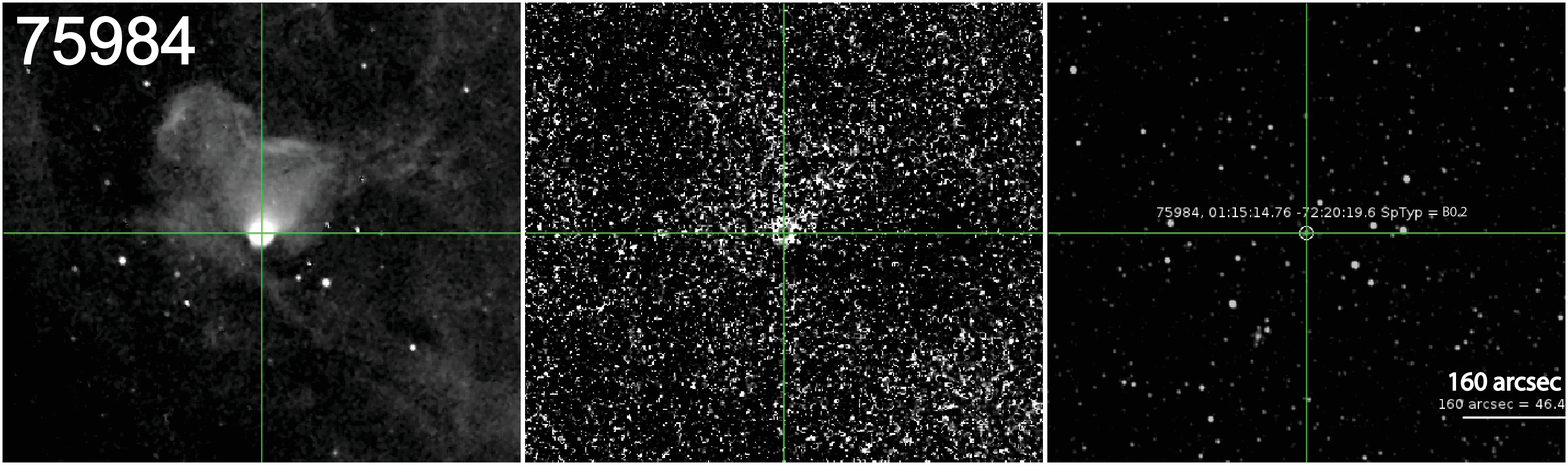}
\caption{
--- continued.}
\end{figure*}

\begin{figure*}[!h]
\plotone{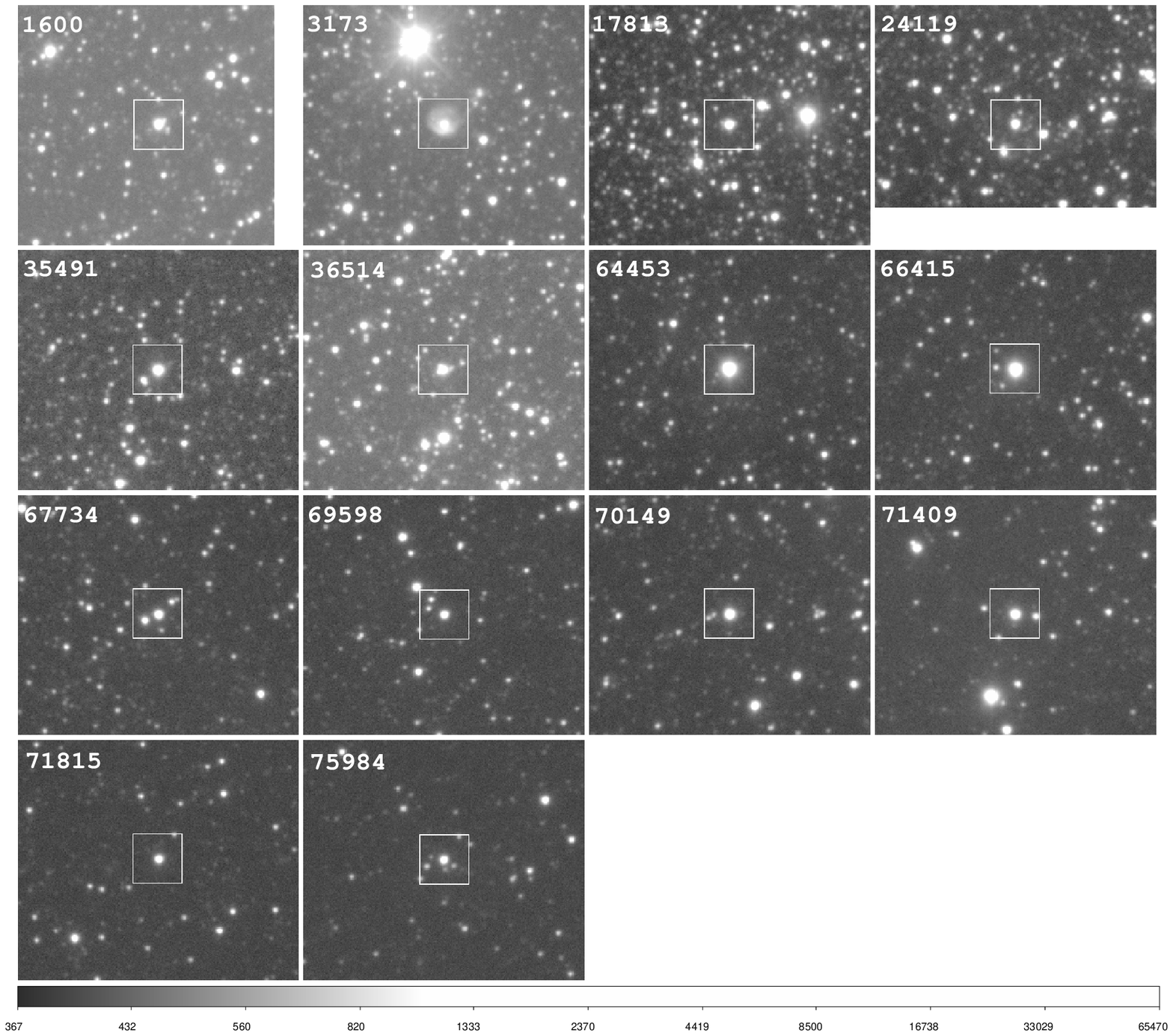}
\caption{
OGLE $I$-band images of the sample objects.
The target stars are centered within 12$\arcsec$ boxes (3.4 pc).  N is up, E to the left.
\label{f_broadband}}
\end{figure*}

\begin{figure*}
\includegraphics[width=8cm]{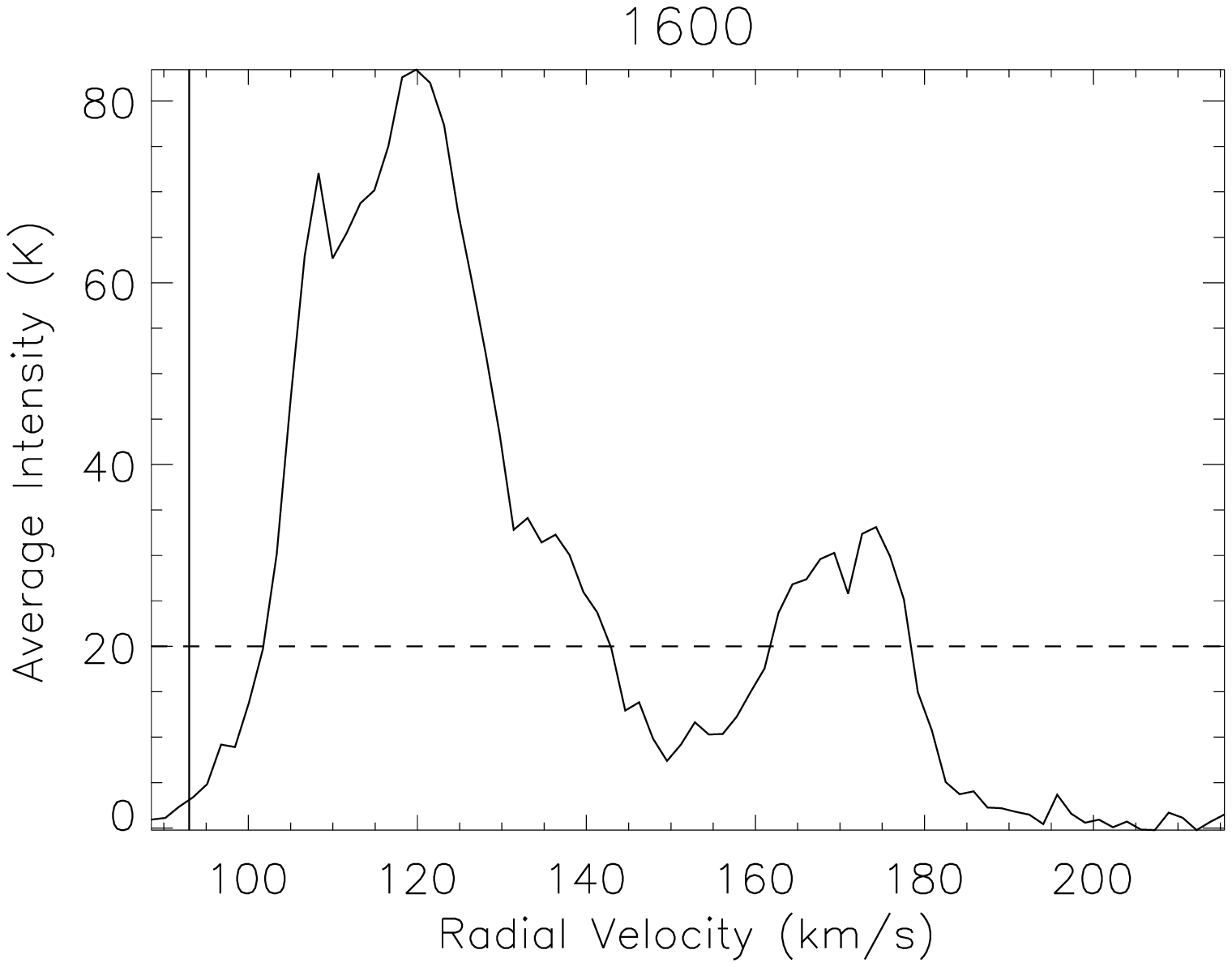}
\includegraphics[width=8cm]{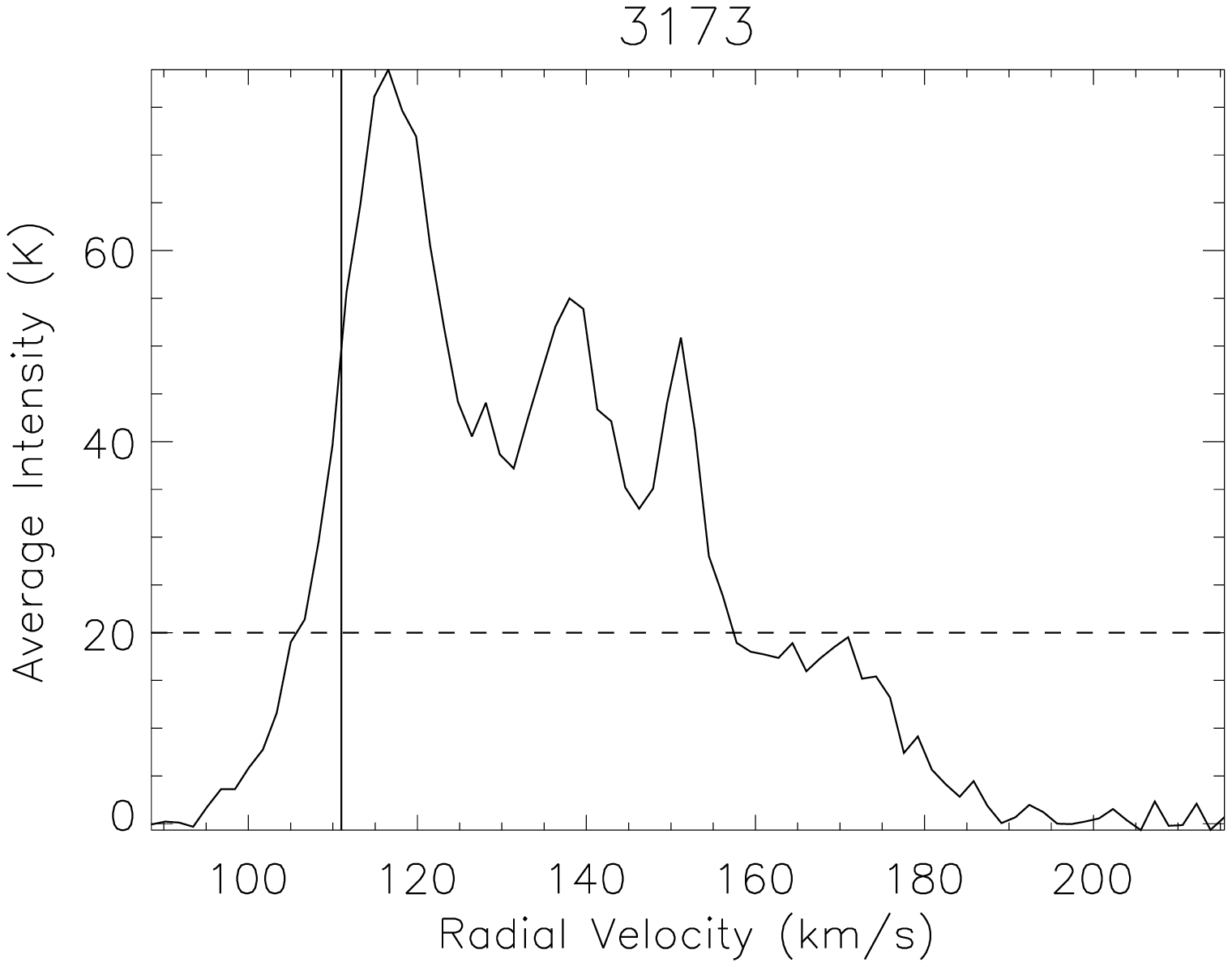}
\includegraphics[width=8cm]{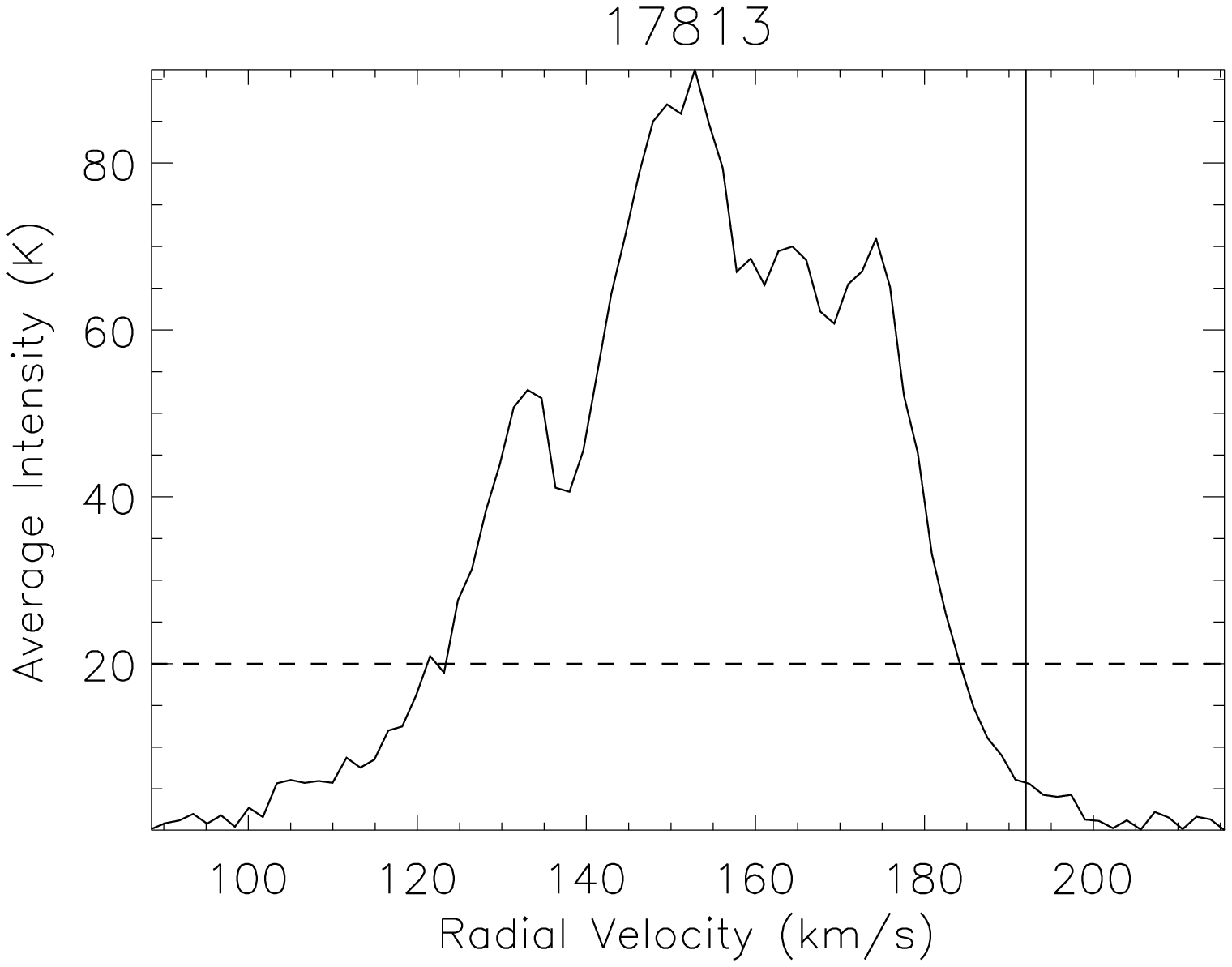}
\includegraphics[width=8cm]{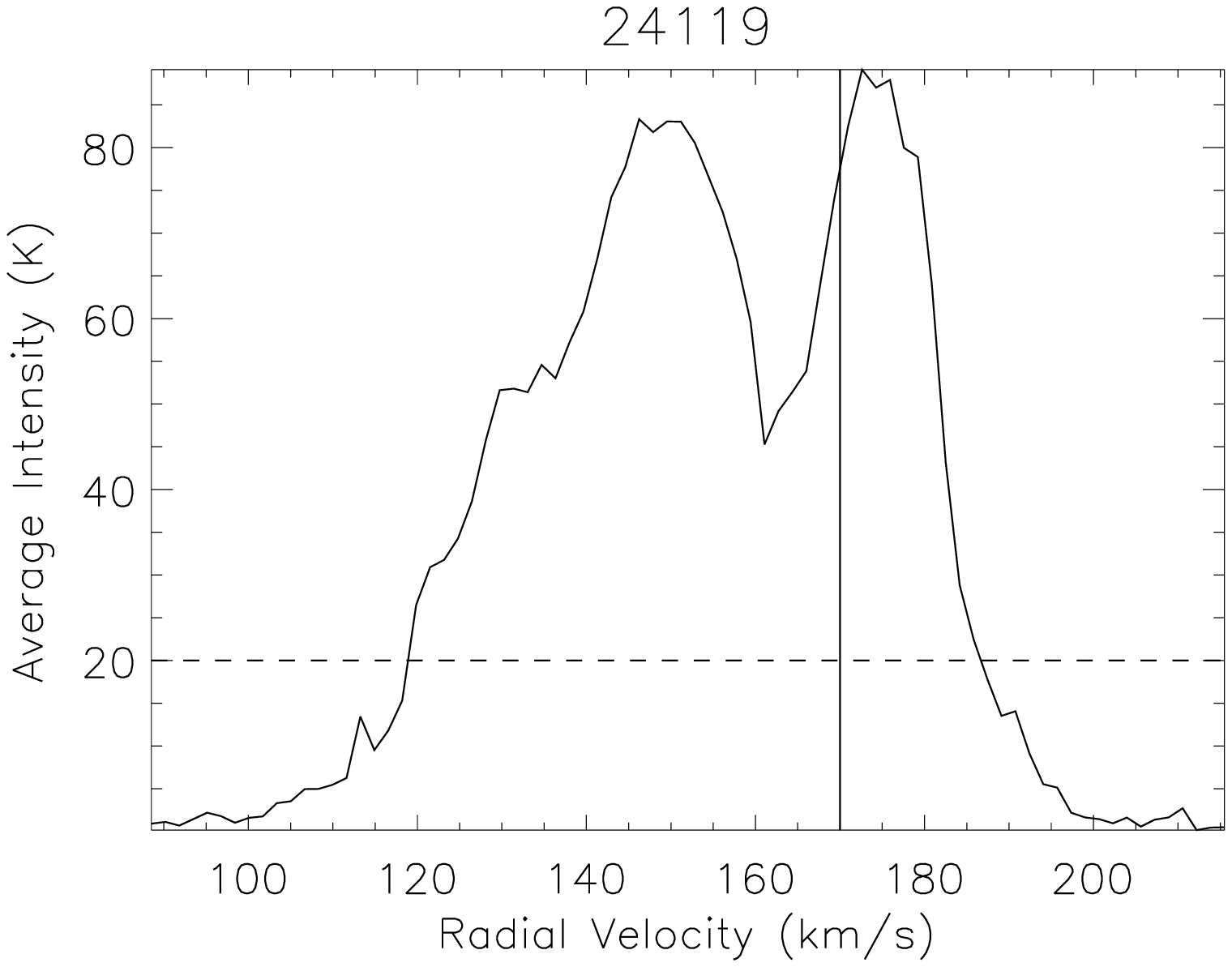}
\includegraphics[width=8cm]{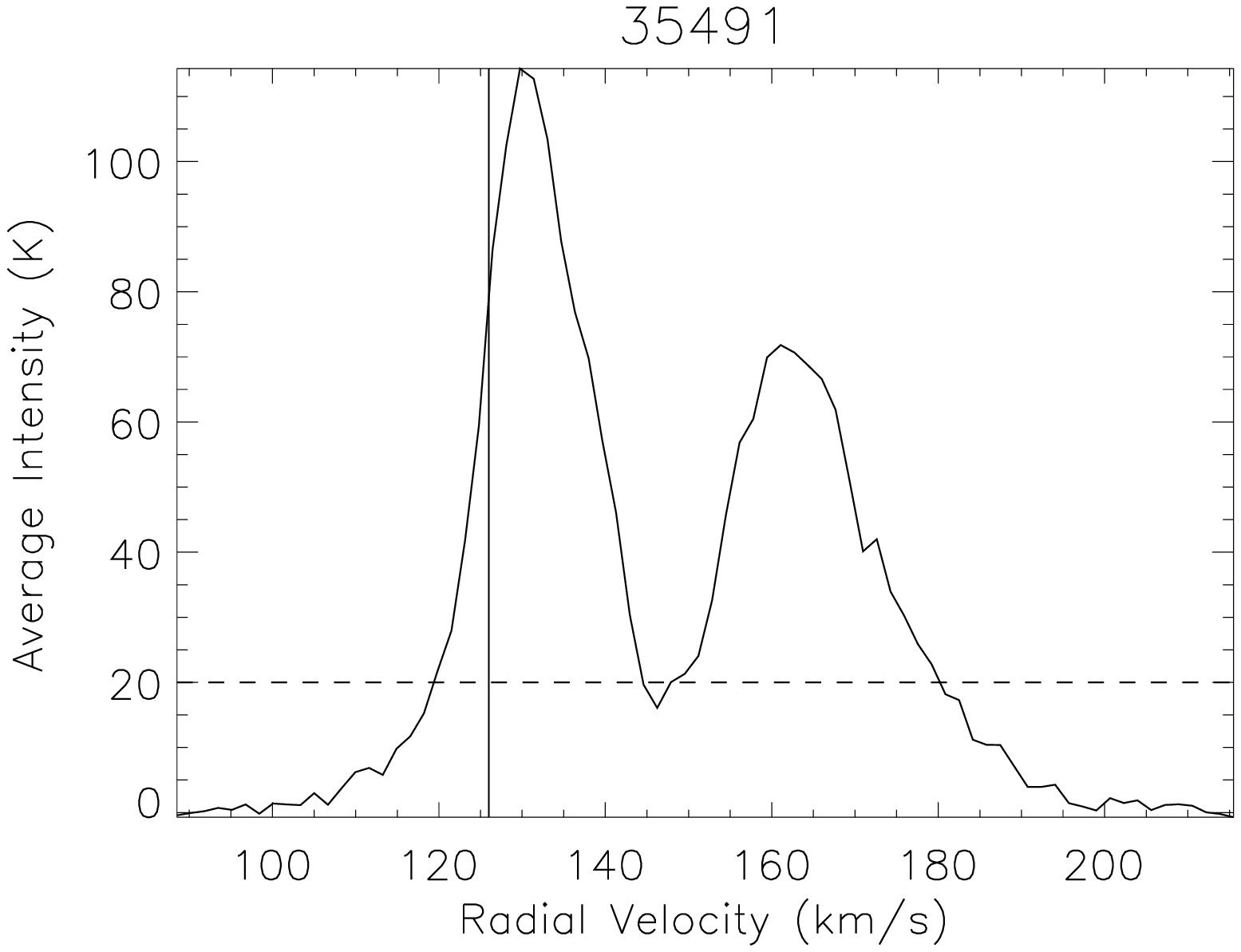}
\includegraphics[width=8cm]{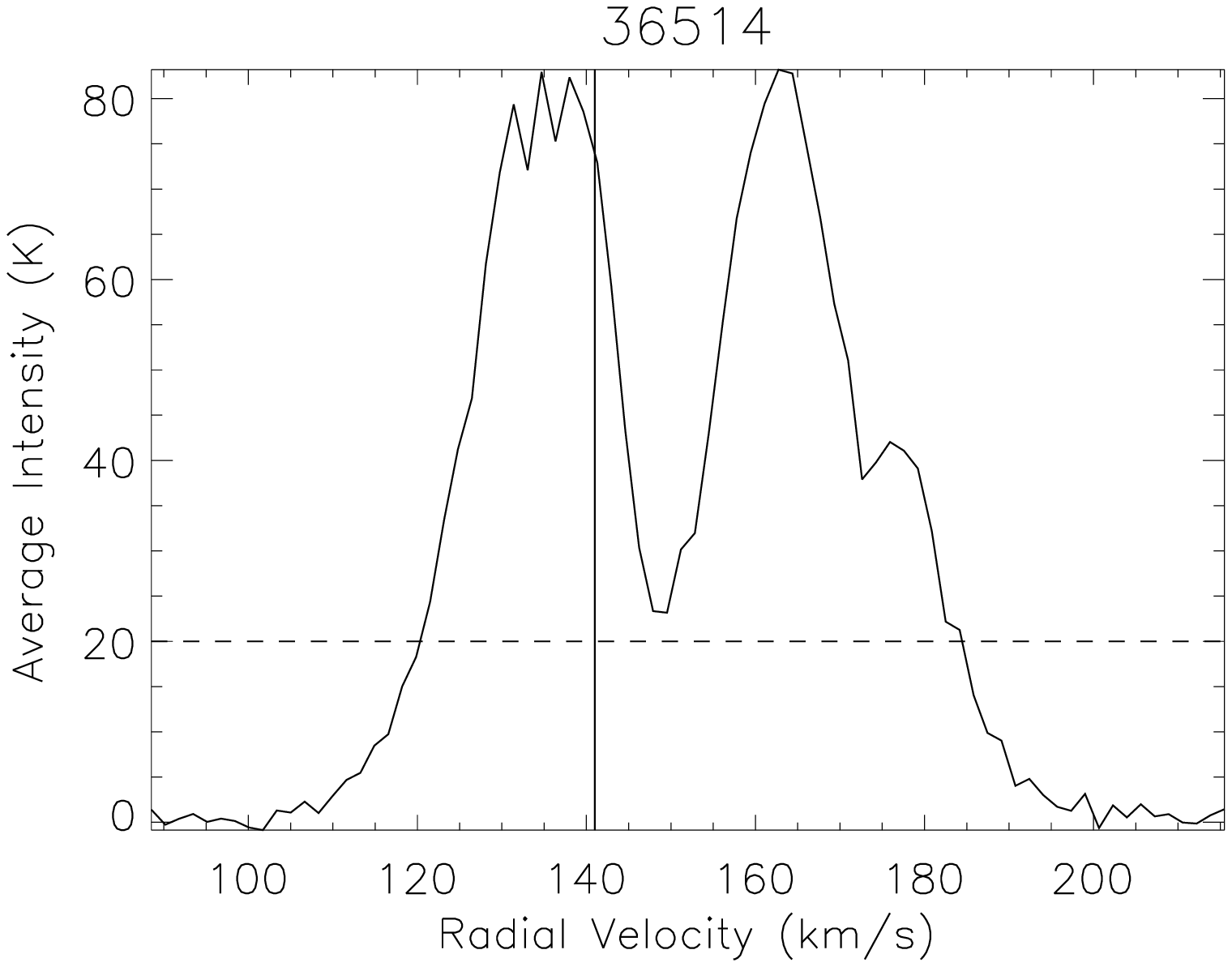}
\includegraphics[width=8cm]{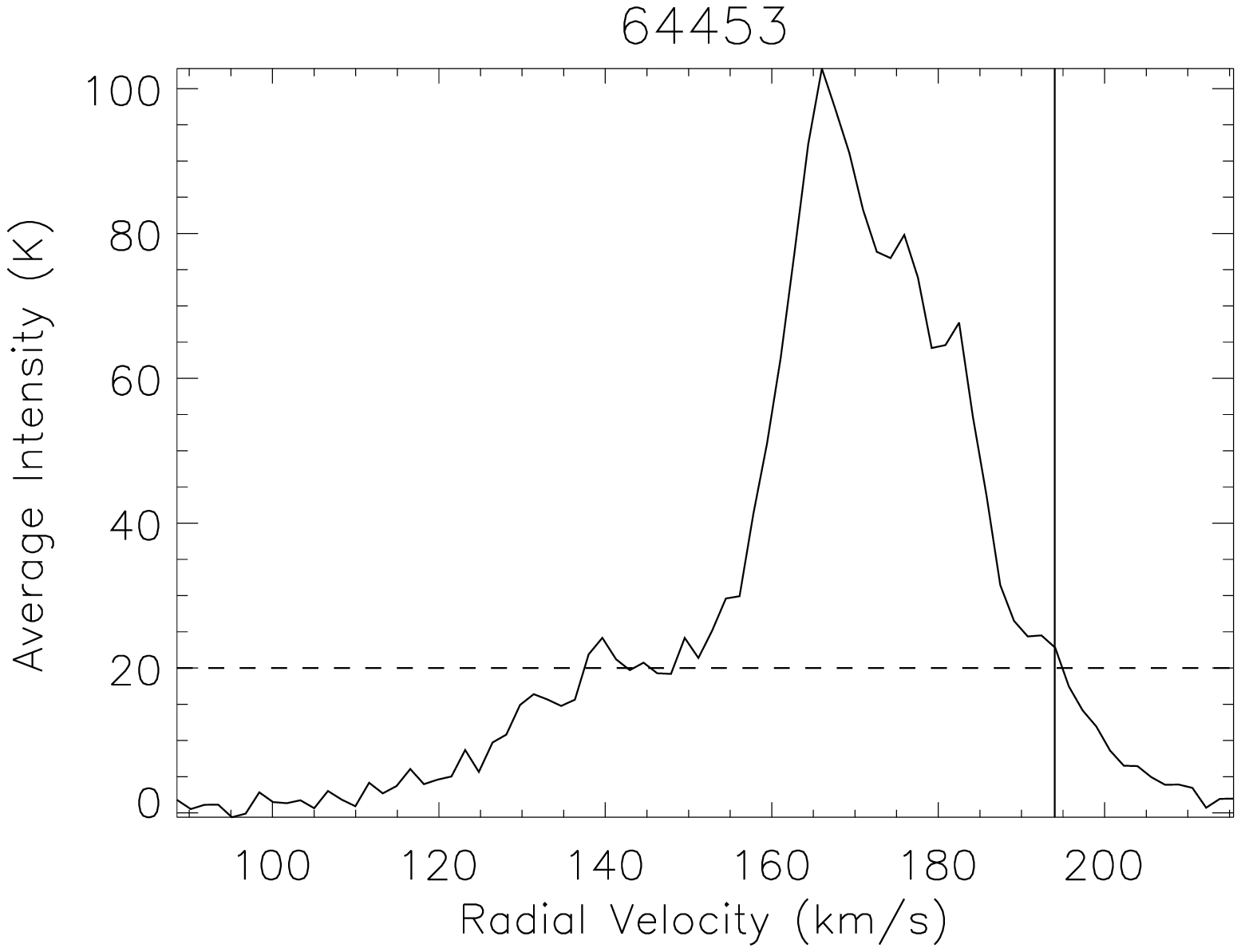}
\hspace*{0.7in}
\includegraphics[width=8cm]{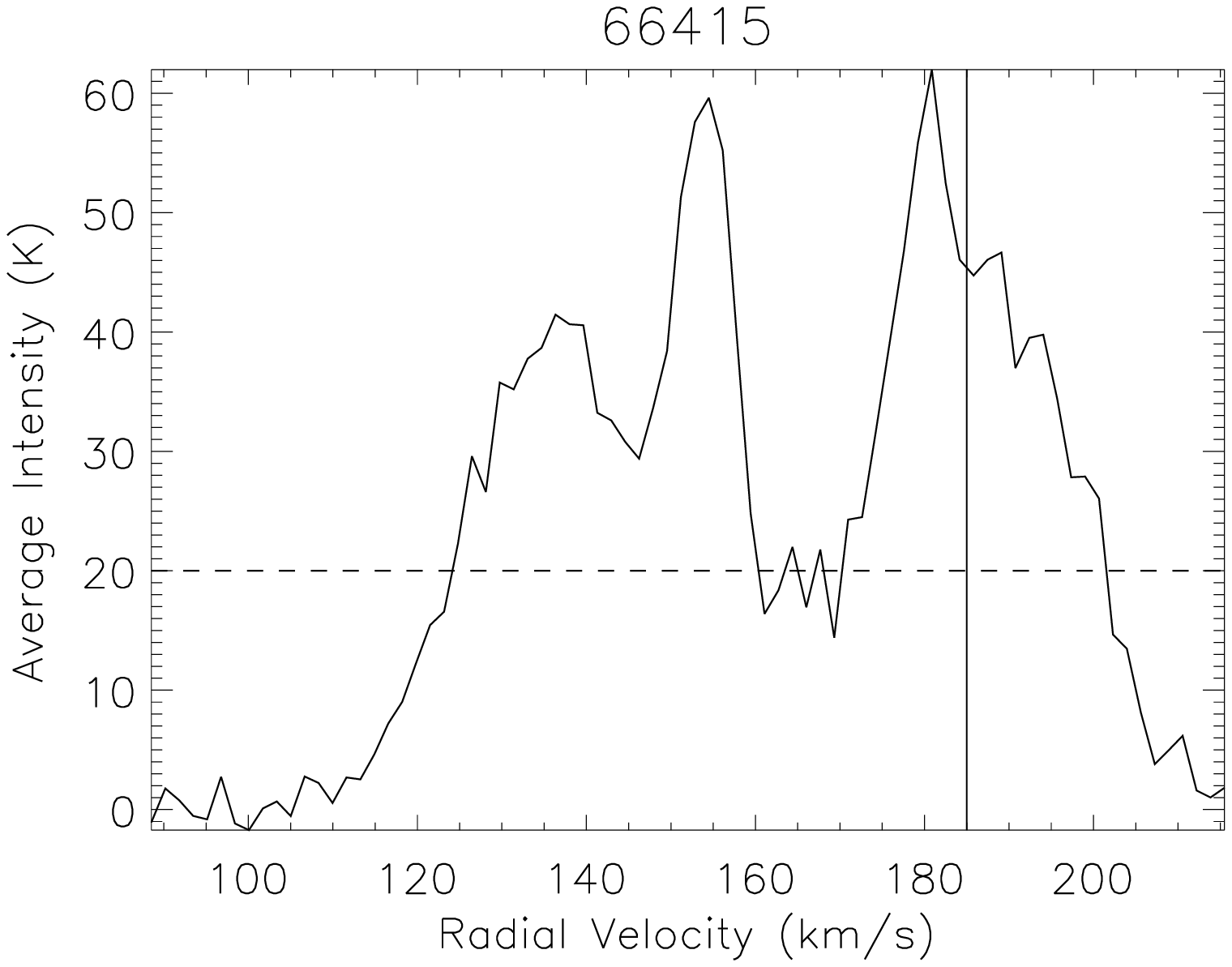}
\caption{The \hi\ velocity structure within a $1\arcmin\times
  1\arcmin$ aperture centered on each target star from the \hi\ survey
  by Stanimirovi\'c et al. (1999), plotted as brightness temperature
  vs radial velocity.  The solid vertical line shows the star's
  measured RV, and the dashed lines shows a nominal criterion for
  significant \hi\ (see text).  For star 75984, note that 
 \hi\ is only observed for velocities $< 215\ \kms$, and so the
 abrupt drop-off at the high-velocity end is likely an artifact.
\label{f_rvhi}}
\end{figure*}

\setcounter{figure}{2}
\begin{figure*}
\includegraphics[width=8cm]{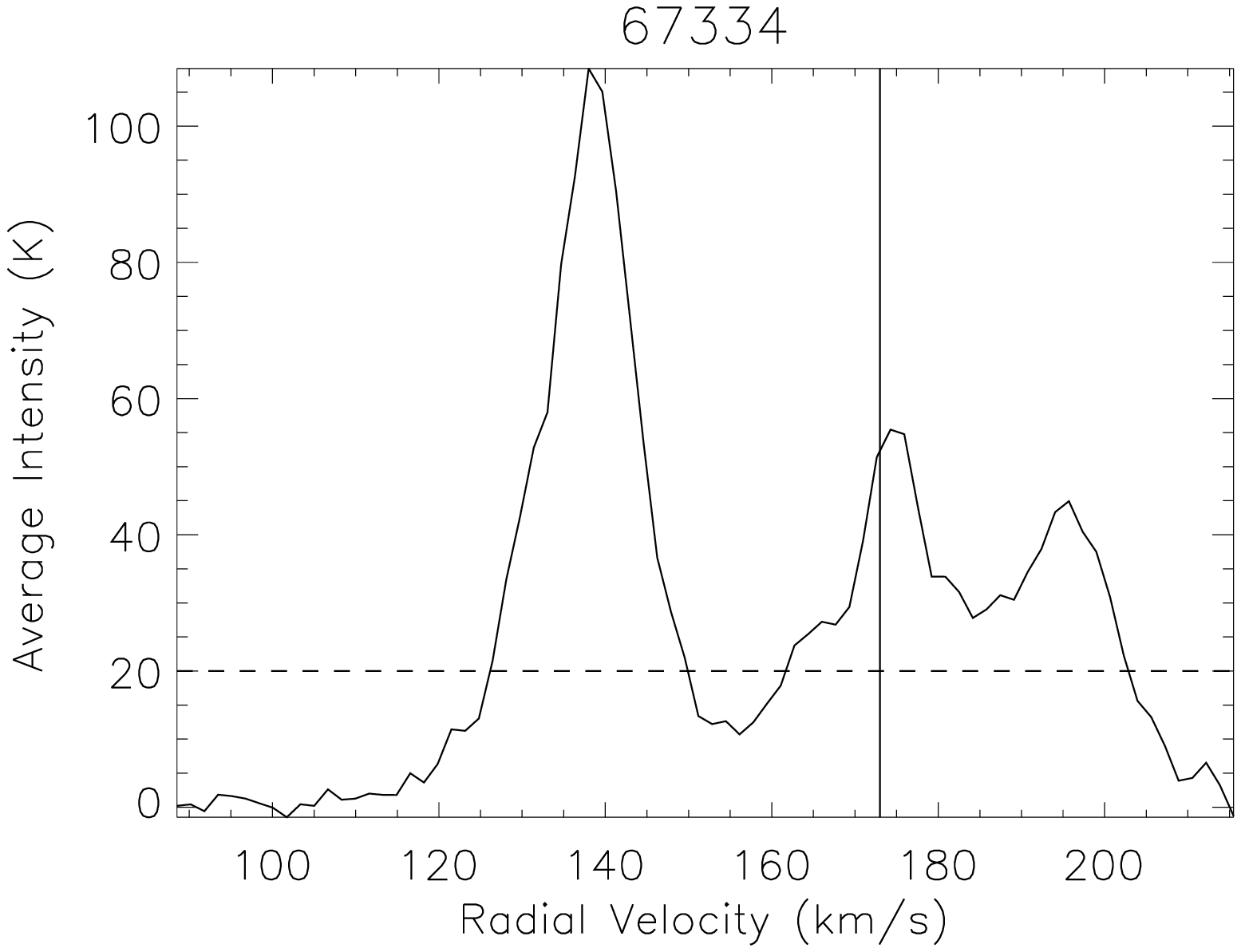}
\includegraphics[width=8cm]{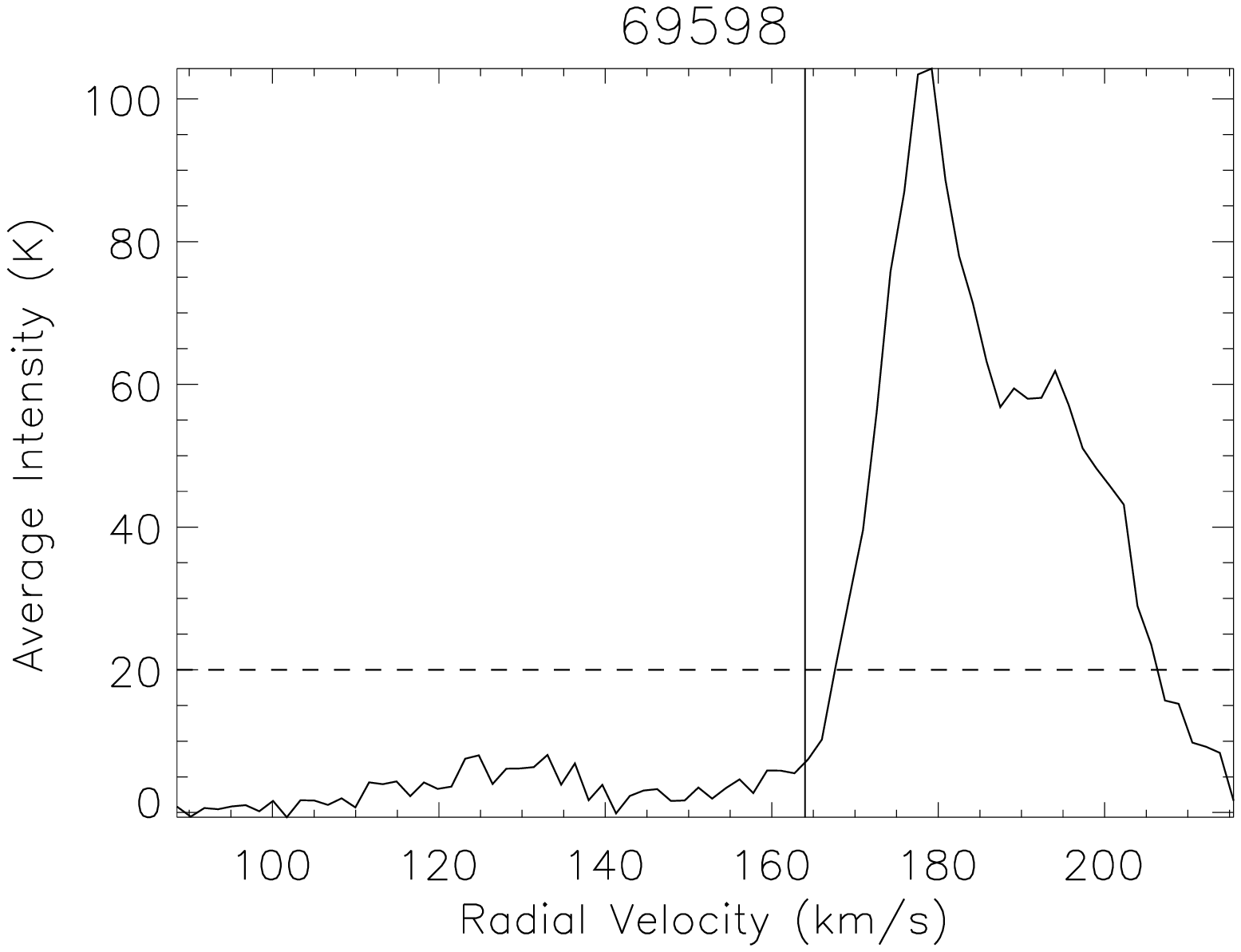}
\includegraphics[width=8cm]{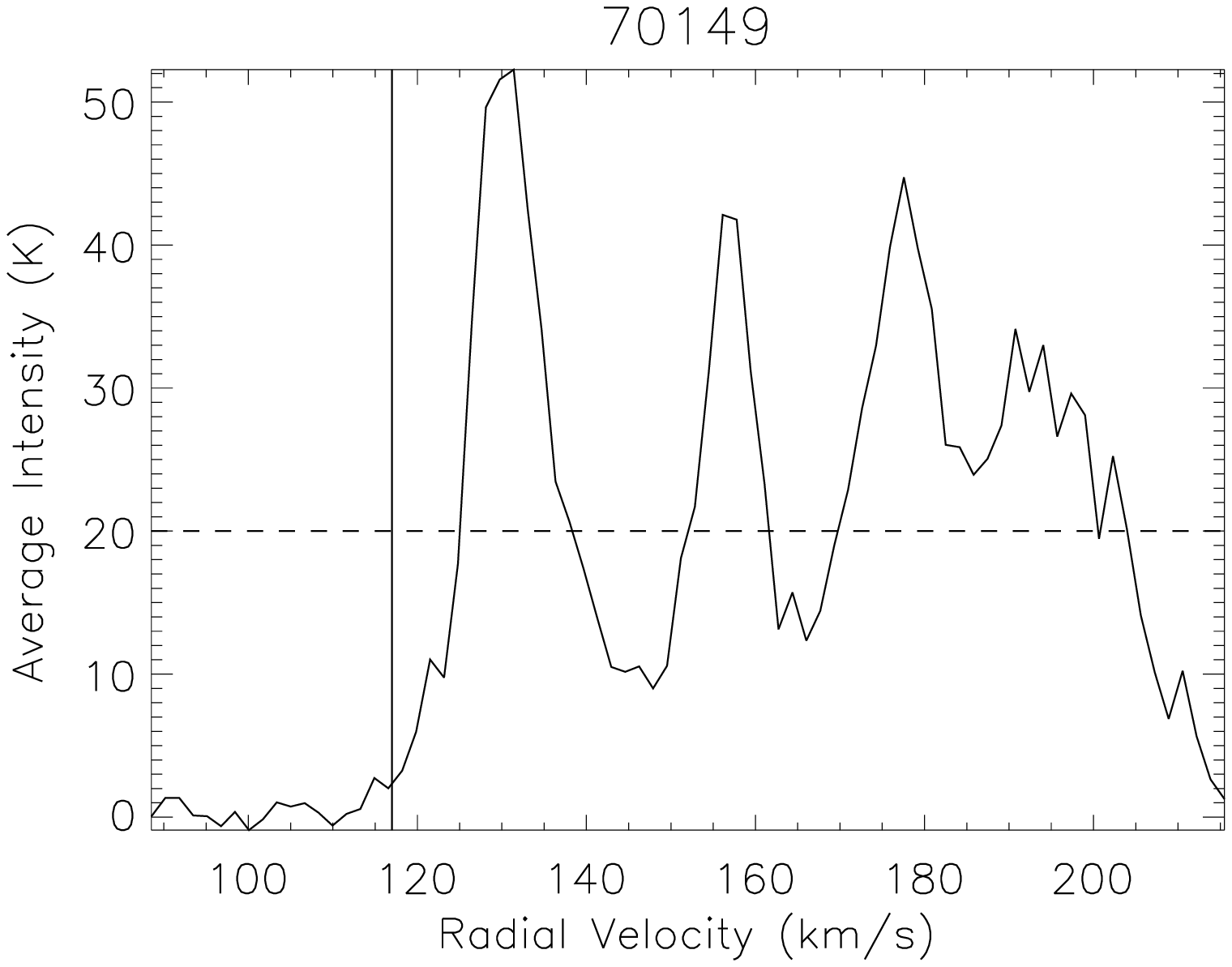}
\includegraphics[width=8cm]{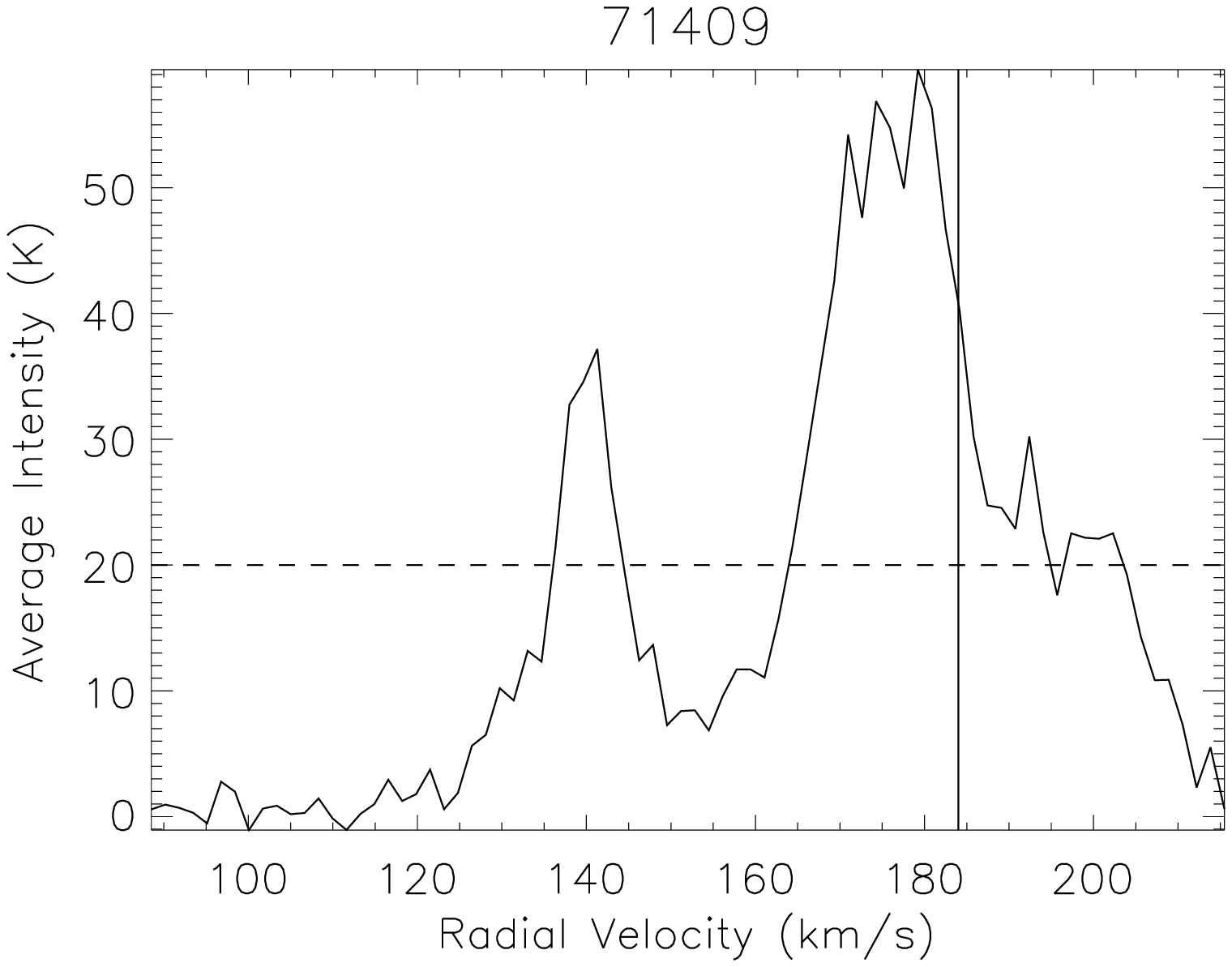}
\includegraphics[width=8cm]{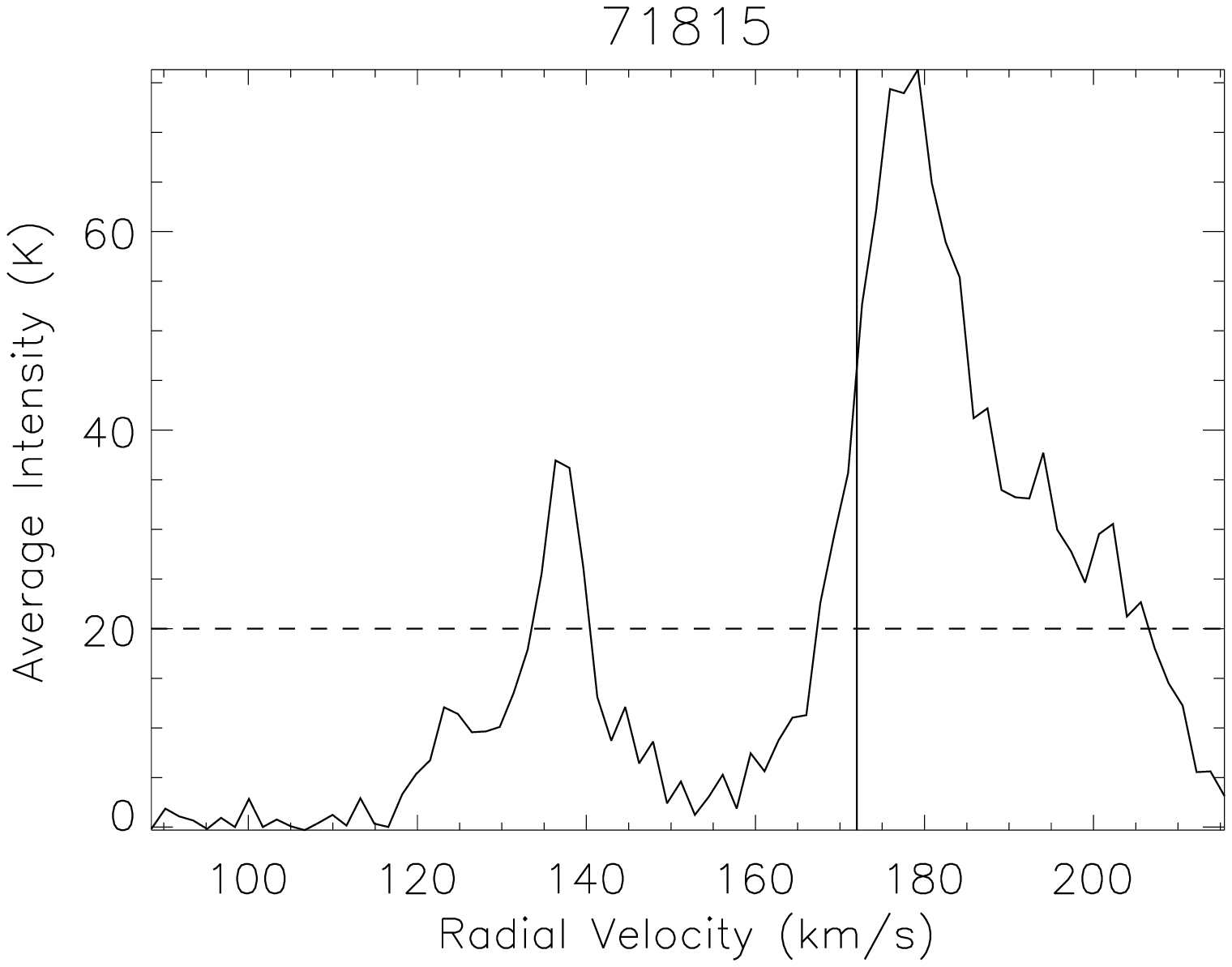}
\hspace*{0.7in}
\includegraphics[width=8cm]{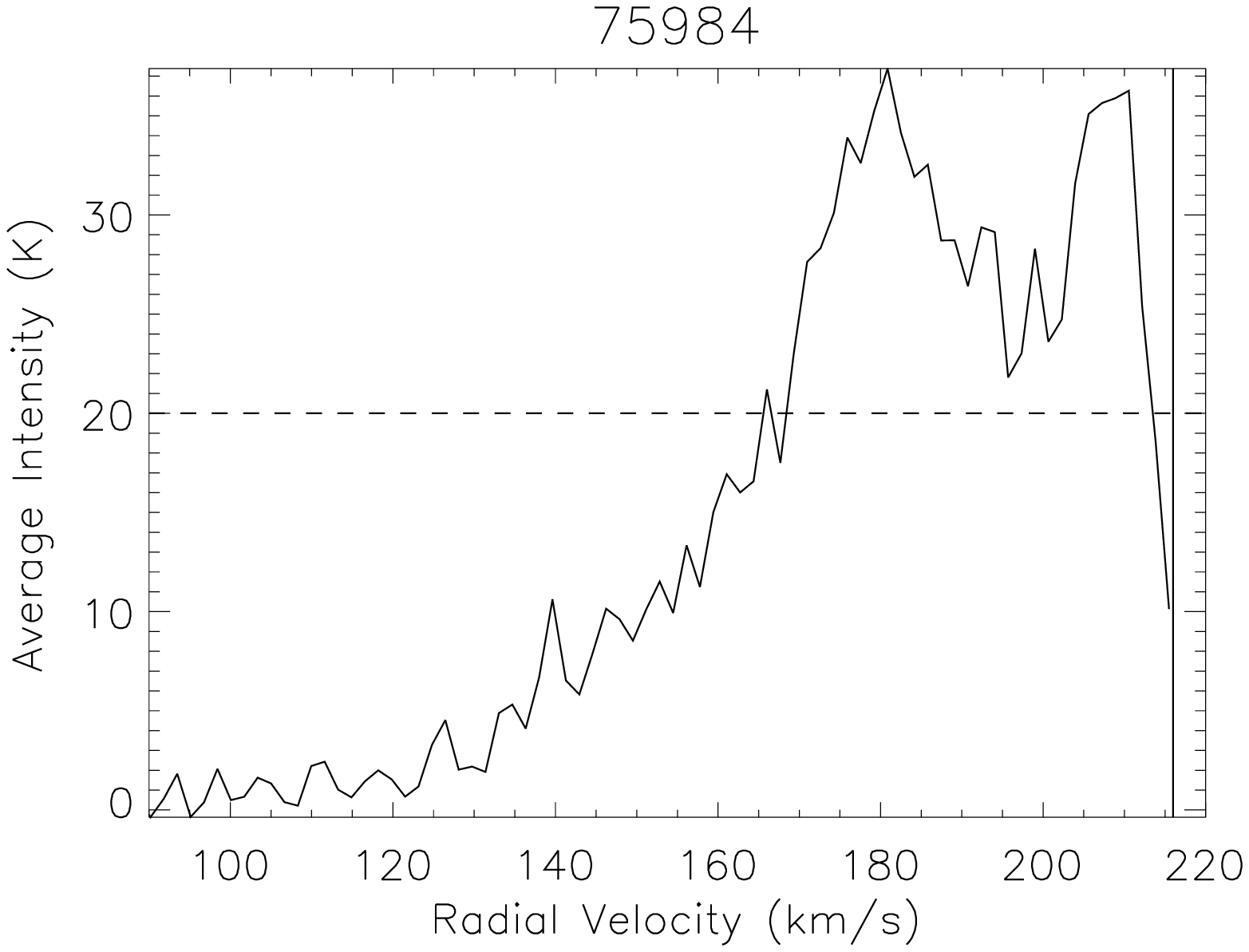}
\caption{
--- continued.}
\end{figure*}

Figure~\ref{f_sample} shows MCELS imaging of the sample, and
Figure~\ref{f_broadband} shows OGLE $I$-band images.  The nebular
images in Figure~\ref{f_sample} are
constructed following Pellegrini et al. (2012).  The left panel for
each object shows \Ha, and the middle panel shows the \sii/\oiii\
ratio image.  The right panel shows the MCELS green continuum band
with an effective wavelength of 5130 \AA\ and $\Delta\lambda=155$ \AA.
Figure~\ref{f_sample} shows that all of these field stars are
significantly isolated and have symmetric nebulae that are difficult
to explain if the stars are runaways ejected from clusters.
The nebulae show regular, spherical morphology with a lower-ionization
envelope traced by \sii\ enclosing the high-ionization nebular center
traced by \oiii.  This Str\"omgren-sphere ionization structure
indicates that the nebulae are most likely optically thick 
(Pellegrini et al. 2012).  Owing to their later spectral type in the
early B-range, stars 17813, 64453, 67334, and 75984 do not generate significant
radiation in \oiii\ and therefore do not show ionization
stratification in Figure~\ref{f_sample}.

Figure~\ref{f_sample} shows that the target stars are centered within
these Str\"omgren spheres.  The H recombination time $t_r = 1/n_e
\alpha_{\rm H}$, where the recombination coefficient $\alpha_{\rm H} =
4\times 10^{-13}\ \rm cm^3\ s^{-1}$ for electron temperature $T_e
= 10^4$ K.  For $n_e=10\ \rm cm^{-3}$, $t_r \sim 10^4$ yr; a runaway
star moving at 100 $\kms$ will travel $\sim$1 pc during this time.
Thus, in such a dense medium, a runaway star will not
noticeably ``outrun'' its nebula, and its Str\"omgren sphere 
will accompany its motion.  However, in the diffuse ISM,
$n_e$ is much lower, increasing $t_r$ proportionately.  For a
high-density value of $n_e=1\ \rm cm^{-3}$, $t_r\sim 10^5$ yr, and a
100 $\kms$ runaway star will travel 10 pc.  The Str\"omgren radius
$R_s \sim 30$ pc for an O9.5 V star, and so
in this case, the star will appear significantly
off-center, in a nebula that is elongated
along the direction of motion.  Stars traveling with 
such high speeds should also show bow shocks.  If the star is within
the diffuse, ionized medium, then $n_e$ is even lower, on the order of
0.1 cm$^{-3}$, with an even longer-lived nebula.

Furthermore, runaway stars with velocities strongly dominated by the
line-of-sight component will remain close to their parent cluster in
projection.  As shown in Figure~\ref{f_broadband}, this is not seen for
our sample objects.  In any case, for this sample of stars, the radial
velocities (RV) shown in Table~\ref{t_sample} are selected to be within 10 $\kms$ of
the nearest \hi\ components having brightness temperature $\geq 20$ K, as
mapped by Stanimirovi\'c et al. (1999).  This is a
nominal criterion for the presence of significant \hi\ in the line of
sight used by Lamb et al. (2013b, in preparation).  Figure~\ref{f_rvhi}
shows that the stellar RVs show excellent correspondence with the
\hi\ velocity components, consistent with the scenario that these
stars formed near their currently observed locations.  Four objects,
1600, 17813, 69598, and 70149,
fall outside the \hi\ threshold criterion and might be candidate
runaway stars.  However, as we shall see below, other criteria suggest
that most of these are also in-situ field stars.
 
We also note that some nebulae, in particular those for 66415 and 69598, show
morphology near their periphery suggesting photo-evaporation by the stellar UV
radiation (\Ha\ images in Figure~\ref{f_sample}).  Such features are common in
star-forming regions, and in extreme form correspond to ``elephant
trunk'' structures pointing radially at the star.  The presence of
large-scale photo-evaporated morphology again strongly suggests that
the surrounding nebulae are indeed the natal clouds for the enclosed O stars.

\section{Degree of Isolation}

We investigate the degree of isolation for these stars more
quantitatively using data from the OGLE-III survey (Udalski et al. 2008) and
a friends-of-friends algorithm (Battinelli 1991; Lamb et
al. 2010).  We first determine the clustering length in a
$18\arcmin\times 9\arcmin$ field within which the target star is
found, using all stars brighter than a completeness limit of 20.5 mag
in $I$.  The clustering length $l$ is the projected distance that 
maximizes the number of clusters consisting of 3 or more stars 
found within $l$ of a cluster member.  Most of the fields show
$l\sim 1 - 2$ pc, depending on mean stellar density.  Since we are
only interested in companion stars that are physically associated with the
target, we distinguish between main sequence stars, which we assume to
be associated, and other stars, which we assume to be background or
foreground stars.  We thus apply a conservative color criterion of $V-I <
0.5$, which discriminates between main sequence stars and Galactic
contamination, which becomes significant at $V-I > 0.5$, according to the Besan\c{c}on stellar
population synthesis models for the Galaxy (Robin et al. 2003).
Table~\ref{t_sample} gives $l$ for each target star, the total
number of companion stars $N$(tot) identified by the friends-of-friends
algorithm, and the number of companions $N$(MS) considered to be main sequence
stars, and therefore physically associated with the target OB
star.  We list the identified companion stars themselves, with their
photometry, in Table~\ref{t_clustering}.  The last column of
Table~\ref{t_clustering} indicates which of these companions are
identified as main sequence stars.

More than one-third of the field OB stars (5 out of 14) show no identified
cluster at the completeness depth of the survey field.  This is a
conservative frequency, since the friends-of-friends clustering
identification includes contaminating background stars.  Furthermore,
small-number statistics will also introduce a few spurious cluster
detections. 
As noted above, four stars have RVs falling outside those of the
line-of-sight \hi\ components.  Of these, 1600 and 69598 show 2 and 6
associated companion stars, respectively (Table~\ref{t_sample}).  It
is therefore extremely likely that 69598 formed in situ with this
sparse group.  The other two stars, 17813 and 70149, show no evidence
of companion stars.

We follow the sparse cluster analysis by Lamb et al. (2010),
examining the ratio between the mass $m_2$ of the second-highest mass
star to that of the target star $m_1$ for each identified group of
companions.  The target masses are from Lamb et al. (2013), and are
simply based on their photometry and spectral types.  We obtain masses
for the companion stars using the statistical method from Lamb et
al. (2013), using OGLE III photometry.  This is done
by generating a probability distribution  
for the observed $VI$ photometry based on the photometric error and
the measured extinction from Zaritsky et al. (2002).  The extinction
$A_V$ for each star is obtained by averaging $A_V$ for all stars
within a 1 arcmin radius, and the adopted uncertainty is the standard
deviation of these extinction values.  For
each companion star, we generate $10^4$ independent realizations of its
$V$ magnitude, $I$ magnitude, and extinction by selecting a random value
from a Gaussian distribution centered on their measured values.  These
realizations are compared with evolutionary tracks at SMC metallicity
from Charbonnel et al. (1993) to obtain a stellar mass for each
realization.  Any realizations existing outside the model parameter space
are discarded as unphysical.  The remaining realizations are averaged
to obtain our adopted mass for each companion star.  Further details
on this technique can be found in Lamb et al. (2013).
Table~\ref{t_sample} lists the target star masses and the mass ratios
$m_2/m_1$, while Table~\ref{t_clustering} gives mass estimates for all
the identified companion stars.

\begin{deluxetable*}{cccccccc}
\scriptsize
\tablewidth{0pt}
\tablecolumns{8}
\tablecaption{Companion stars}
\tablehead{
\colhead{RA and Dec (J2000)\tablenotemark{a}} & \colhead{$V$\tablenotemark{a}} & 
\colhead{$I$\tablenotemark{a}} & \colhead{$V$ err\tablenotemark{a}} & \colhead{$I$ err\tablenotemark{a}} &
\colhead{$A_V$\tablenotemark{b}} & \colhead{$m(\rm M_\odot)$} & \colhead{MS\tablenotemark{c}} }
\startdata
\cutinhead{1600}
0:42:10.43 ~ --73:13:58.2 & 19.786 & 19.855 & 0.119 & 0.200 & 0.43 & 2.3 & Y \\
0:42:09.44 ~ --73:13:57.9 & 19.821 & 18.850 & 0.098 & 0.081 & 0.41 & 1.9 & N \\
0:42:09.62 ~ --73:13:55.9 & 19.075 & 18.535 & 0.128 & 0.102 & 0.38 & 2.5 & Y \\
\cutinhead{3173}
0:43:36.83 ~ --73:02:28.9 & 18.197 & 18.824 & 0.094 & 0.112 & 0.36 & 4.7 & Y \\
0:43:36.97 ~ --73:02:28.3 & \nodata & 18.949 & \nodata & 0.130 & 0.38 & \nodata & \nodata \\
0:43:37.28 ~ --73:02:25.5 & 19.976 & 19.514 & 0.248 & 0.177 & 0.37 & 2.0 & Y \\
0:43:36.28 ~ --73:02:28.4 & 18.438 & 19.119 & 0.103 & 0.129 & 0.37 & 4.1 & Y \\
0:43:36.53 ~ --73:02:28.0 & 17.686 & 18.611 & 0.065 & 0.106 & 0.37 & \nodata & \nodata \\
\cutinhead{24119}
0:52:38.30 ~ --73:26:15.2 & 19.372 & 19.342 & 0.325 & 0.190 & 0.25 & 2.4 & Y \\
0:52:37.39 ~ --73:26:17.7 & 20.138 & 19.827 & 0.115 & 0.203 & 0.26 & 1.9 & Y \\
0:52:37.66 ~ --73:26:18.4 & 19.847 & 19.763 & 0.097 & 0.175 & 0.26 & 2.2 & Y \\
0:52:37.92 ~ --73:26:14.4 & 19.609 & 18.843 & 0.085 & 0.074 & 0.26 & 2.3 & N \\
\cutinhead{36514}
0:56:17.80 ~ --72:17:30.1 & 20.561 & 20.504 & 0.157 & 0.316 & 0.32 & 1.8 & Y \\
0:56:17.99 ~ --72:17:27.2 & 20.881 & 20.953 & 0.194 & 0.427 & 0.32 & 1.7 & Y \\
0:56:18.25 ~ --72:17:31.4 & \nodata & 21.096 & \nodata & 0.363 & 0.33 & \nodata & \nodata\\
0:56:18.28 ~ --72:17:24.0 & 19.835 & 18.801 & 0.060 & 0.060 & 0.33 & 1.7 & N \\
0:56:18.59 ~ --72:17:28.7 & 21.298 & 21.121 & 0.186 & 0.370 & 0.33 & 1.5 & Y \\
0:56:18.69 ~ --72:17:26.0 & 19.642 & 18.656 & 0.063 & 0.056 & 0.33 & 1.8 & N \\
0:56:18.75 ~ --72:17:33.1 & 19.560 & 19.556 & 0.048 & 0.121 & 0.35 & 2.5 & Y \\
0:56:18.96 ~ --72:17:33.6 & 19.987 & 19.566 & 0.068 & 0.126 & 0.35 & 2.0 & Y \\
0:56:18.98 ~ --72:17:28.0 & 20.402 & 20.198 & 0.108 & 0.195 & 0.32 & 1.9 & Y \\
0:56:19.43 ~ --72:17:26.5 & 21.330 & 20.687 & 0.214 & 0.279 & 0.33 & 1.5 & Y \\
0:56:16.30 ~ --72:17:27.2 & 19.528 & 18.752 & 0.056 & 0.056 & 0.34 & 2.3 & N \\
0:56:16.88 ~ --72:17:28.3 & \nodata & 19.022 & \nodata & 0.085 & 0.33 & \nodata & \nodata \\
\cutinhead{64453}
1:06:40.96 ~ --73:10:25.9 & 19.611 & 19.228 & 0.646 & 0.375 & 0.29 & 2.2 & Y \\
1:06:41.15 ~ --73:10:27.6 & 20.057 & 20.235 & 0.218 & 0.271 & 0.31 & 2.1 & Y \\
\cutinhead{66415}
1:07:41.13 ~ --72:51:04.0 & 19.504 & 18.746 & 0.077 & 0.065 & 0.25 & 2.4 & N \\
1:07:41.12 ~ --72:50:51.6 & 20.128 & 20.207 & 0.073 & 0.182 & 0.26 & 2.2 & Y \\
1:07:41.47 ~ --72:50:59.4 & 18.653 & 18.002 & 0.056 & 0.034 & 0.26 & 3.0 & N \\
1:07:41.45 ~ --72:50:55.4 & 19.253 & 19.331 & 0.037 & 0.098 & 0.27 & 2.9 & Y \\
\cutinhead{67334}
1:08:08.79 ~ --72:38:21.1 & 17.668 & 16.396 & 0.019 & 0.010 & 0.41 & 2.4 & N \\
1:08:07.05 ~ --72:38:16.1 & 20.026 & 19.069 & 0.063 & 0.075 & 0.41 & 1.9 & N \\
1:08:07.33 ~ --72:38:16.6 & 18.663 & 17.537 & 0.032 & 0.019 & 0.42 & 2.0 & N \\
1:08:07.90 ~ --72:38:23.5 & 18.806 & 18.835 & 0.032 & 0.065 & 0.40 & 3.1 & Y \\
\cutinhead{69598}
1:09:27.27 ~ --72:01:21.3 & 20.841 & 19.955 & 0.154 & 0.151 & 0.22 & 1.6 & N \\
1:09:27.48 ~ --72:01:23.2 & 17.497 & 17.673 & 0.013 & 0.023 & 0.22 & 5.7 & Y \\
1:09:27.61 ~ --72:01:25.4 & 17.919 & 18.105 & 0.014 & 0.033 & 0.22 & 5.1 & Y \\
1:09:28.10 ~ --72:01:30.6 & 18.639 & 18.724 & 0.023 & 0.055 & 0.21 & 3.6 & Y \\
1:09:28.11 ~ --72:01:40.9 & 16.094 & 16.283 & 0.008 & 0.010 & 0.23 & 8.9 & Y \\
1:09:28.26 ~ --72:01:20.3 & 16.728 & 15.741 & 0.009 & 0.008 & 0.20 & 4.9 & N \\
1:09:28.44 ~ --72:01:26.5 & 20.750 & 19.859 & 0.125 & 0.147 & 0.19 & 1.6 & N \\
1:09:28.82 ~ --72:01:39.8 & 20.453 & 19.684 & 0.092 & 0.117 & 0.22 & 1.8 & N \\
1:09:28.99 ~ --72:01:33.9 & 18.617 & 18.712 & 0.018 & 0.056 & 0.22 & 3.7 & Y \\
1:09:29.19 ~ --72:01:43.3 & 20.620 & 19.637 & 0.146 & 0.112 & 0.22 & 1.6 & N \\
1:09:29.54 ~ --72:01:33.4 & 19.143 & 19.242 & 0.036 & 0.086 & 0.20 & 3.2 & Y 
\enddata
\label{t_clustering}
\end{deluxetable*}

\setcounter{table}{2}
\begin{deluxetable*}{cccccccc}
\scriptsize
\tablewidth{0pt}
\tablecolumns{8}
\tablecaption{Companion stars (continued)}
\tablehead{
\colhead{RA and Dec (J2000)\tablenotemark{a}} & \colhead{$V$\tablenotemark{a}} & 
\colhead{$I$\tablenotemark{a}} & \colhead{$V$ err\tablenotemark{a}} & \colhead{$I$ err\tablenotemark{a}} &
\colhead{$A_V$\tablenotemark{b}} & \colhead{$m(\rm M_\odot)$} & \colhead{MS\tablenotemark{c}} }
\startdata
\cutinhead{70149}
1:09:49.13 ~ --72:30:19.5 & 19.701 & 18.789 & 0.074 & 0.066 & 0.31 & 2.0 & N \\
1:09:49.40 ~ --72:30:20.3 & 19.400 & 18.506 & 0.048 & 0.048 & 0.32 & 2.1 & N \\
1:09:48.27 ~ --72:30:24.0 & \nodata & 20.349 & \nodata & 0.214 & 0.31 & \nodata & \nodata \\
\cutinhead{75984}
1:15:15.37 ~ --72:20:18.7 & 20.238 & 19.677 & 0.122 & 0.140 & 0.30 & 1.8 & Y \\
1:15:15.61 ~ --72:20:21.0 & 17.992 & 18.155 & 0.023 & 0.036 & 0.30 & 5.0 & Y \\
1:15:15.95 ~ --72:20:24.0 & 18.501 & 18.640 & 0.024 & 0.050 & 0.36 & 4.0 & Y \\
1:15:13.26 ~ --72:20:15.7 & 19.918 & 18.972 & 0.052 & 0.067 & 0.32 & 1.8 & N \\
1:15:13.62 ~ --72:20:36.2 & 19.130 & 18.189 & 0.035 & 0.035 & 0.37 & 2.3 & N \\
1:15:13.88 ~ --72:20:35.2 & 21.018 & 20.059 & 0.172 & 0.170 & 0.37 & 1.5 & N \\
1:15:13.87 ~ --72:20:13.8 & 18.335 & 18.494 & 0.024 & 0.045 & 0.29 & 4.5 & Y \\
1:15:14.04 ~ --72:20:07.8 & 20.413 & 19.475 & 0.093 & 0.109 & 0.30 & 1.6 & N \\
1:15:14.18 ~ --72:20:20.9 & 18.662 & 18.802 & 0.026 & 0.064 & 0.34 & 3.9 & Y \\
1:15:14.63 ~ --72:20:21.7 & 19.613 & 19.369 & 0.075 & 0.104 & 0.31 & 2.1 & Y \\
1:15:14.71 ~ --72:20:29.8 & 19.398 & 19.486 & 0.048 & 0.112 & 0.40 & 2.8 & Y
\enddata
\tablenotetext{a}{From OGLE III (Udalski et al. 2008).}
\tablenotetext{b}{From Zaritsky et al. (2002).}
\tablenotetext{c}{Indicates whether star is considered to be on the main sequence.}
\end{deluxetable*}

Table~\ref{t_clustering} demonstrates that most of the target field OB
stars do show evidence of companion stars.  If these are
indeed physically associated with the targets, it provides even
stronger evidence that these OB stars formed in situ.  
The mass ratios $m_2/m_1$ may be slightly larger than those
reported by Lamb et al. (2010), whose maximum $m_2/m_1$ is 0.17.
However, that work was based on much deeper
observations with the {\it Hubble Space Telescope}, for a smaller
sample; whereas it is likely that some of our OGLE stars are in fact
unresolved multiples.  We also note our high frequency of
non-detections of companions.  As discussed by Lamb et al. (2010), while their
$m_2/m_1$ ratios are low, they are still expected with reasonable
probabilities of 10 -- 20\% if the IMF and cluster mass functions are
independent.  This contrasts with scenarios in which the maximum-mass
star depends on cluster mass (e.g., Weidner \& Kroupa 2006; Vanbeveren 1982).

\begin{figure}[h]
\includegraphics[width=9cm]{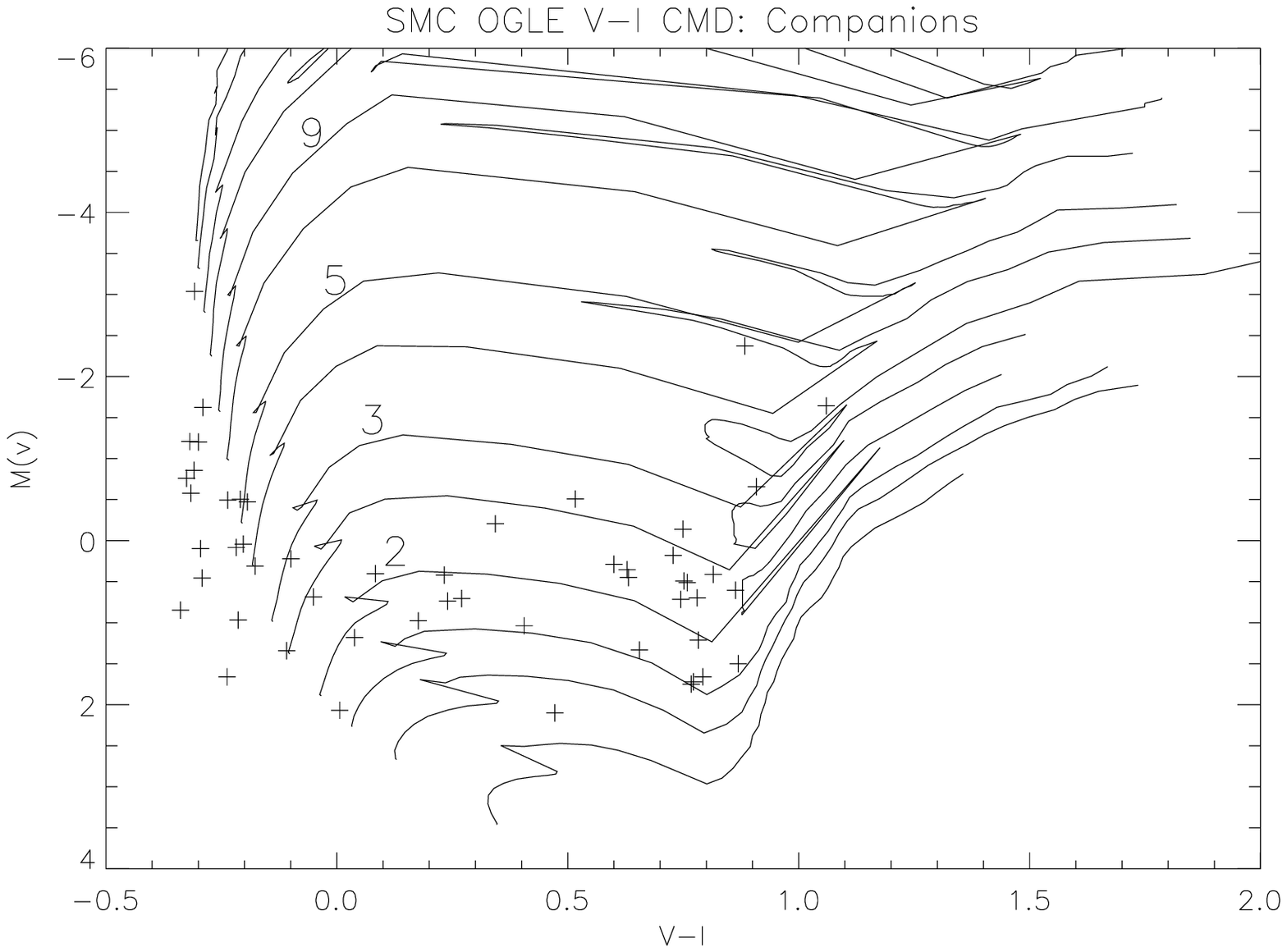}
\caption{Composite color-magnitude diagram of all the stars identified
to be in sparse groups in Table~\ref{t_clustering}, excluding the
target OB stars.  Evolutionary tracks from Charbonnel et al. (1993)
are overlaid, with model stellar masses shown.}
\label{f_cmd}
\end{figure}

Figure~\ref{f_cmd} shows the color-magnitude diagram for all the
identified companion stars in Table~\ref{t_clustering}.
We generate a composite IMF for all the identified clusters, following
the formalism of Scalo (1986) and assuming a single, power-law form,
having slope $\Gamma$ given by, 
\begin{equation}
\Gamma = d \log \xi(\log m)/d \log m \quad ,
\end{equation}
where $\xi(\log m)$ is the mass function given in units of stars born
per logarithmic stellar mass $m(M_\odot$) per unit area (kpc$^2$).
We obtain $\xi(\log m)$
by counting the number of stars per mass bin and dividing by the area
covered by the observations.  We account for the different size
mass bins by normalizing each bin to one dex in mass.
When constructing the IMF, our star counts include only the companion
stars identified as main sequence stars above (Table~\ref{t_clustering}).

\begin{figure}[h]
\includegraphics[width=9cm]{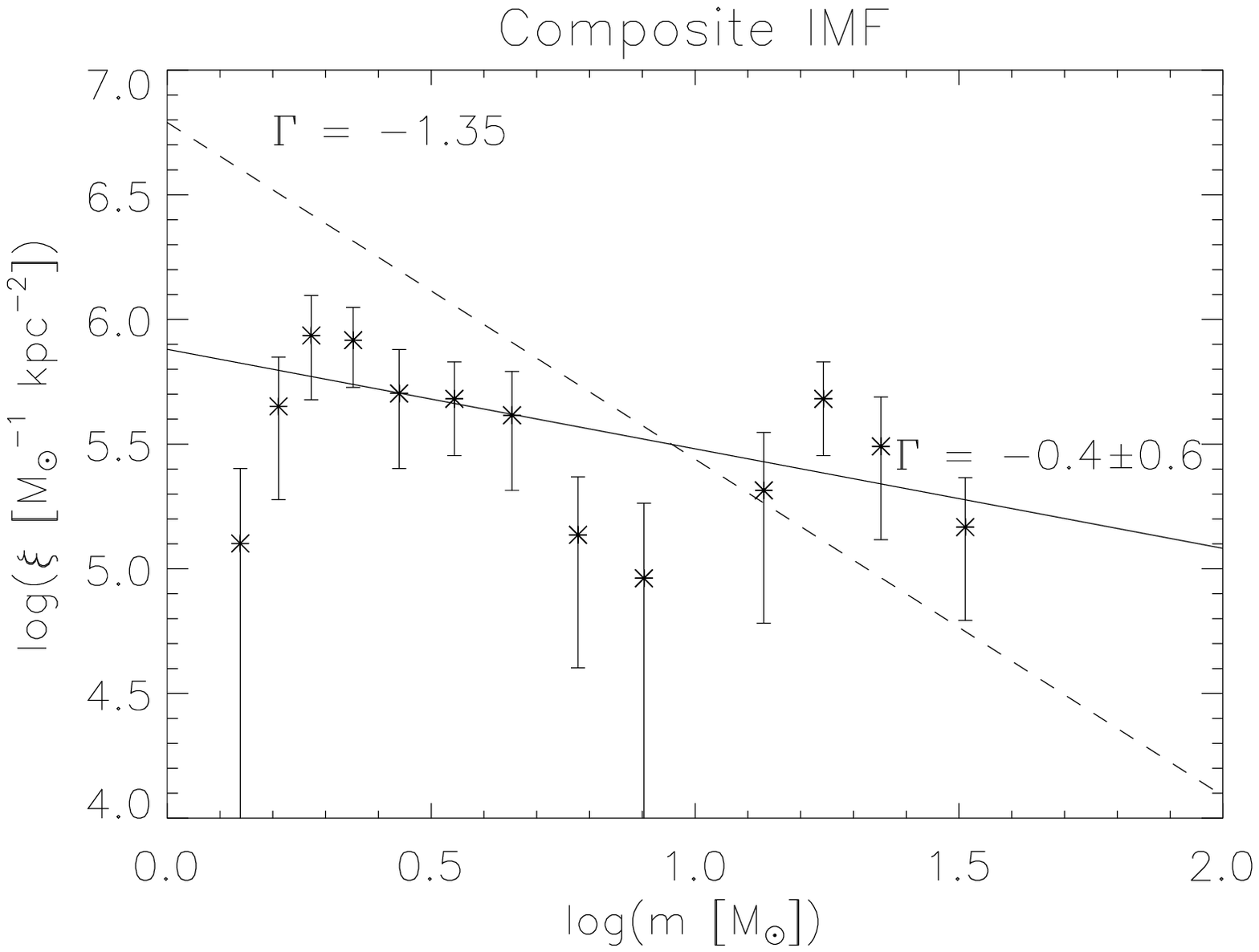}
\caption{Composite IMF of all the sparse groups in
  Table~\ref{t_clustering}, including the target OB stars.  The fitted
slope is shown with a solid line, while the dashed line shows the
Salpeter value. Root-$N$ error bars are shown.}
\label{f_imfall}
\end{figure}

\begin{figure}[h]
\includegraphics[width=9cm]{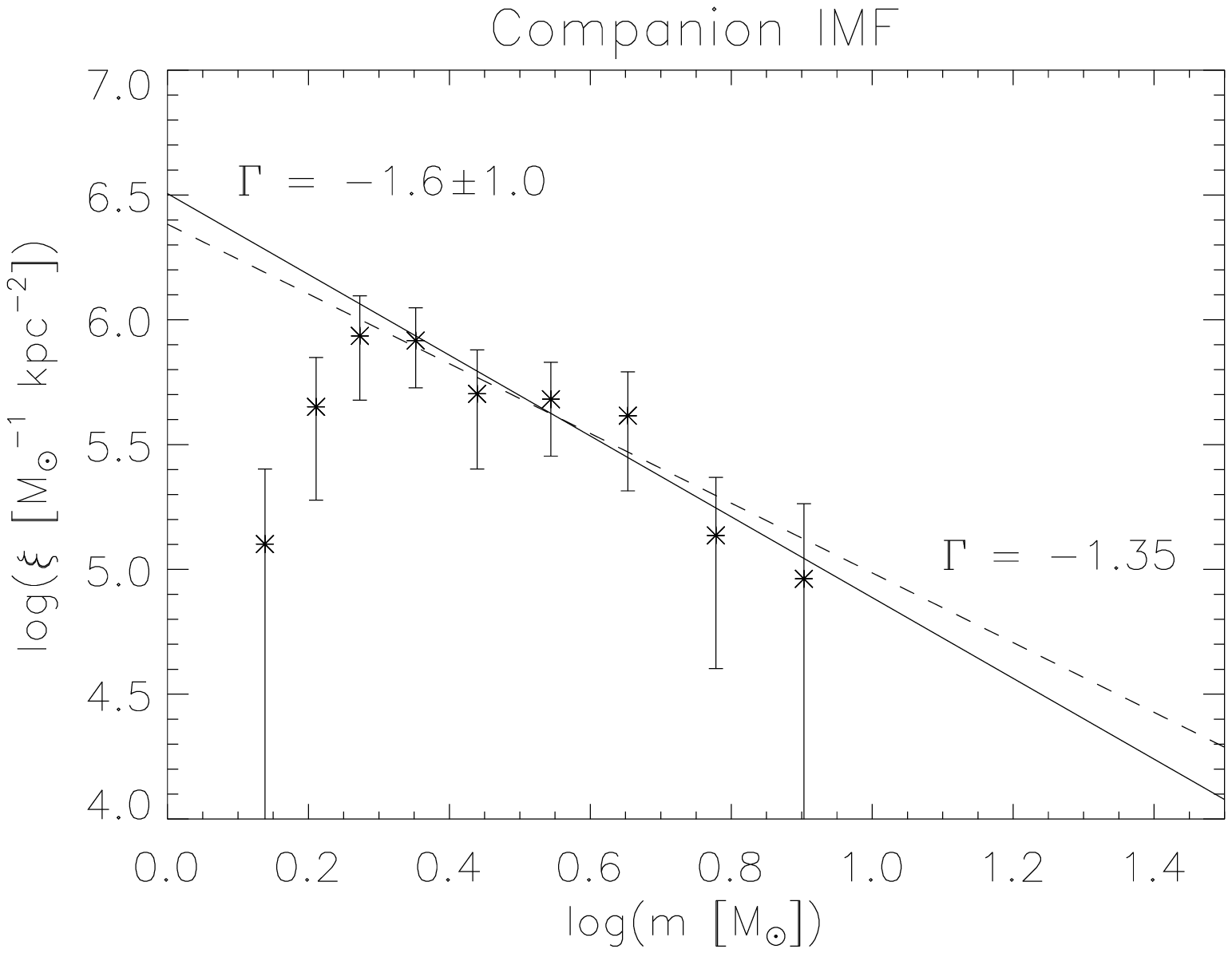}
\caption{Same as Figure~\ref{f_imfall}, but excluding the target OB stars.}
\label{f_imfcompanions}
\end{figure}

Figure~\ref{f_imfall} shows the composite mass function for all the
target stars with companions.  Since the sample selection criteria
are based on field OB star status, the composite IMF 
must be much flatter than the usual
Salpeter power-law slope of $\Gamma = -1.35$.
Figure~\ref{f_imfall} shows that this is indeed the case, with the
composite IMF slope of $\Gamma = -0.4 \pm 0.6$.  This is similar to the
composite slope for companions found by Lamb et al. (2010) of $\Gamma
= 0.1\pm 1.0$.  Excluding the target stars, the composite IMF slope is
$\Gamma = -1.6\pm 1.0$, which turns out to be consistent with the Salpeter value
(Figure~\ref{f_imfcompanions}). 

Finally, we note that five of the 14 target stars, 17813, 35491,
70149, 71409, and 71815, show no evidence of associated companions
down to the completeness limit of $I = 20.5$ (Table~\ref{t_sample}). 
We further examined these fields for detected stars fainter than the
completeness limit, and only one target, 17813, shows an additional
detected companion fainter than this threshold.
If these five OB stars formed in situ, as is suggested by our
fairly rigorous selection criteria, then they set even stronger limits
to the star formation density than the rest of the sample.  This
includes the possibility that they formed in complete isolation.
The stars 17813 and 70149 are also among the objects whose RV
falls outside the observed \hi\ kinematic components
(Figure~\ref{f_rvhi}), and thus these objects have
the highest probability of being runaway stars.  However, the
difference in velocities is only by 8 and 9 $\kms$, respectively, and as
argued earlier, the spherical symmetry of the nebulae, MCELS-S73 and
MCELS-S174, is hard to explain if the stars are runaways
(Figure~\ref{f_sample}).  The nebular densities 
are also fully commensurate with the normal
densities of star-forming \hii\ regions (Table~\ref{t_nebulae}).

We have not addressed the possibility that the target field OB stars
are binaries.  Given that the majority of OB stars in clusters are
binaries (e.g., Sana et al. 2012), it is reasonable to expect that
many, if not most of the field OB stars are binaries as well.  To
date, none of the stars in our sample are known binary stars. 
Whether or not any turn out to be binaries will also offer vital diagnostics of
star formation models, for example, whether the binary frequency is
different among field OB stars relative to that in clusters.  

\section{Conclusion}

In summary, we present a sample of 14 isolated OB stars that appear to have
formed in the field, under extremely sparse star-forming conditions.
The stars are selected to be at least 28 pc from any other OB star,
and they are centered within fairly circular, symmetric \hii\ regions,
with no evidence of bow shocks, implying that they are not runaway stars with
transverse velocities.  They are also not line-of-sight runaways,
based on their radial velocities relative to local \hi, and the fact
that no obvious parent clusters are near the line of sight.
\hii\ regions generated by two of the stars show evidence of photo-evaporated
structures, as seen in other star-forming regions.
We carried out a friends-of-friends analysis of the stellar fields for
our sample, which shows that most of these OB stars have $\lesssim 7$
main sequence companions in projection.  The presence of companions,
if they are indeed physically associated, further implies that the
target stars formed in situ.  However, more than 
one third of the sample (5 out of 14 stars) shows no companions above the
completeness limit, and it remains possible that these OB stars formed
in complete isolation.

Our sample of stars presents some of the strongest evidence to date that
massive OB stars can and do form in relative isolation.  Their
existence poses strong empirical constraints on theories of star
formation and challenges proposed relations between cluster mass and
maximum stellar mass.  We also stress that while this sample
represents strong selection criteria, there are many more objects that
may similarly have formed in situ.

\acknowledgments

This work was supported by NSF grants AST-0907758 and AST-0806476.
C.T.K. was supported by the Undergraduate
Research Opportunities Program at the University of Michigan.
We thank Anne Jaskot for assistance with the MCELS data.

\end{document}